\documentclass[universe,review,accept,pdftex,oneauthor]{Definitions/mdpi} 

\firstpage{1} 
\pubvolume{10}
\issuenum{8}
\articlenumber{312}
\pubyear{2024}
\copyrightyear{2024}
\externaleditor{Academic Editors:  Roman Pasechnik and Wolfgang Schafer}
\datereceived{26 June 2024} 
\daterevised{25 July 2024} 
\dateaccepted{28 July 2024} 
\datepublished{30 July 2024} 
\hreflink{https:// doi.org/10.3390/universe10080312} 

\Title{Axion-like Particle Effects on Photon Polarization in High-Energy Astrophysics}

\TitleCitation{Axion-like Particle Effects on Photon Polarization in High-Energy Astrophysics}

\Author{{Giorgio Galanti} 
 \orcidA{}}

\AuthorNames{Giorgio Galanti}

\AuthorCitation{Galanti, G.}

\address[1]{%
{{INAF}
, Istituto di Astrofisica Spaziale e Fisica Cosmica di Milano, Via A. Corti 12, {20133} 
 Milano, Italy; gam.galanti@gmail.com}}

\abstract{In this review, we present a self-contained introduction to axion-like particles (ALPs) with a particular focus on their effects on photon polarization: both theoretical and phenomenological aspects are discussed. We derive the photon survival probability in the presence of photon--ALP interaction, the corresponding final photon degree of linear polarization, and the polarization angle in a wide energy interval. The presented results can be tested by current and planned missions such as IXPE (already operative), eXTP, XL-Calibur, NGXP, XPP in the X-ray band and like COSI (approved to launch), e-ASTROGAM, and AMEGO in the high-energy range. Specifically, we describe ALP-induced polarization effects on several astrophysical sources, such as galaxy clusters, blazars, and gamma-ray bursts, and we discuss their real detectability. In particular, galaxy clusters appear as very good observational targets in this respect. Moreover, in the very-high-energy (VHE) band, we discuss a peculiar ALP signature in photon polarization, in principle capable of proving the ALP existence. Unfortunately, present technologies cannot detect photon polarization up to such high energies, but the observational capability of the latter ALP signature in the VHE band could represent an interesting challenge for the future. As a matter of fact, the aim of this review is to show new ways to make progress in the physics of ALPs, thanks to their effects on photon polarization, a topic that has aroused less interest in the past, but which is now timely with the advent of many new polarimetric missions.}


\begin{document}

\section{Introduction}

The Standard Model (SM) of particle physics provides a satisfactory theoretical baseline able to explain all known processes acting on elementary
particles. However, the SM cannot be considered as the ultimate theory of fundamental interactions since it does not really unify the electroweak and the strong interaction and does not contain gravity. Moreover, the SM describes neither dark matter nor dark energy, which together make up the majority of our Universe. Therefore, the SM represents a low-energy manifestation of a more fundamental theory describing and unifying all four fundamental interactions at the quantum~level.

Several proposals have been presented with the aim of extending and completing the SM, such as
multidimensional Kaluza--Klein theories~\cite{kaluzaklein1,kaluzaklein2,turok}, four-dimensional unsymmetrical models~\cite{susy1,susy2,susy3,susy4,susy5}, and superstring and superbrane theories~\cite{string1,string2,string3,string4,string5,axiverse,abk2010,cicoli2012}. A remarkable fact is that all these theories invariably predict the existence of axion-like particles (ALPs, for a review, see, e.g.,~\cite{alp1,alp2}).

ALPs represent a generalization of the axion, the pseudo-Goldstone boson originating from the breakdown of the global Peccei--Quinn symmetry ${\rm U}(1)_{\rm PQ}$, introduced to solve the strong CP problem (see, e.g.,~\cite{axionrev1,axionrev2,axionrev3,axionrev4}). The axion is characterized by a strict relationship between its mass and its coupling to photons, and it interacts also with gluons and fermions so that the Peccei--Quinn mechanism can work. Instead, two aspects differentiate ALPs from axions: (i) the ALP mass $m_a$ is unrelated to the two-photon-to-ALP coupling $g_{a\gamma\gamma}$, and (ii)~ALPs primarily interact with photons only, while other possible interactions can safely be neglected in the present context. ALPs are currently considered among the best candidates for being the long-sought dark matter~\cite{preskill,abbott,dine,arias2012}. Furthermore, ALPs produce two significant effects in the presence of external magnetic fields: (i) photon--ALP oscillations~\cite{mpz,raffeltstodolsky}, which are similar to the oscillations of massive neutrinos with different flavors with the difference that, in the ALP case, the external magnetic field is necessary for spin compensation between spin-0 ALPs and spin-1 photons, and (ii) the change of the polarization state of \mbox{photons~\cite{mpz,raffeltstodolsky}}. Due to their faint interaction with photons, ALPs can hardly be detected over the limited distances of laboratory experiments. Instead, the astrophysical context allows us to overcome this restriction with photons that can originate at cosmological distances and can cross several magnetized media, where photon--ALP interaction can take place. Therefore, the astrophysical environment appears as the most promising way to study the ALP physics, especially in the high-energy (HE) and in the very-high-energy (VHE) band (see, e.g.,~\cite{irastorzaredondo,grRew} for a review). 

In particular, photon--ALP oscillations occur in different magnetized media, such as inside active galactic nuclei (AGN; see, e.g.,~\cite{trg2015} and Section \ref{sec4.1}). Specifically, photon--ALP oscillations explain the detection of photons with energies of up to $\sim$$400 \, \rm GeV$ originated from flat spectrum radio quasars (FSRQs, an AGN class), while conventional physics prevents any emission for energies above $\sim$$20 \, \rm GeV$ (see also Section \ref{sec4.1}): this fact represents the {\it first hint} at ALP existence~\cite{trgb2012}. Moreover, photon--ALP interaction solves the anomalous redshift dependence of the spectra of BL~Lacs (another AGN class)~\cite{grdb}. In particular, within conventional physics, the values assumed by the BL~Lac VHE-emitted spectral indexes decrease as the redshift grows, while no redshift dependence is expected, since cosmological effects are not present and selection biases are excluded, as discussed in~\cite{grdb}. The introduction of photon--ALP conversion inside the extragalactic space increases the Universe transparency~\cite{drm2007,dgr2011,grExt}, partially avoiding the absorption of VHE photons due to their interaction with the extragalactic background light (EBL, see also Section \ref{sec4.4} \mbox{and~\cite{dgr2013,franceschinirodighiero,gptr}}), and makes BL~Lac VHE-emitted spectral indexes redshift independent. Since this is the only possibility according to the physical expectation, this result represents the {\it {second hint} 
} at ALP existence~\cite{grdb}. Moreover, photon--ALP conversion inside the blazar jet, in the host galaxy, in the extragalactic space and in the Milky Way produces two features~\cite{gtre2019,gtl2020}: (i)~spectral irregularities and (ii) photon excess in blazar spectra at VHE, which can be detected by current and planned observatories, such as ASTRI~\cite{astri}, CTA~\cite{cta}, GAMMA-400~\cite{g400}, HAWC~\cite{hawc}, HERD~\cite{herd}, LHAASO~\cite{LHAASOsens}, and TAIGA-HiSCORE~\cite{desy}. Furthermore, ALP-induced spectral irregularities produced inside the galaxy cluster turbulent magnetic field have been used to place constraints on the ALP parameter space~\cite{fermi2016,CTAfund}. In addition, photon--ALP interaction provokes modifications on stellar evolution~\cite{straniero}; it has been employed inside galaxy clusters as an explanation of the spectral distortions of the continuum thermal emission ($T$$ \sim $$2\, \rm {\rm keV}\text{--}8 \, \rm keV$)~\cite{thermal} and of the unexpected spectral line at $3.55 \, \rm keV$, viewed as dark matter decay into ALPs with their subsequent oscillations into photons~\cite{DMdecay}. ALPs have also been used in order to explain a blazar line-like feature~\cite{wang}. The recent detection of the gamma-ray burst GRB 221009A by the LHAASO Collaboration~\cite{LHAASO,LHAASOspectrumHigh} up to $18 \, \rm TeV$ strongly challenges conventional physics, while the photon--ALP scenario naturally explains the VHE observation of this GRB, which represents the {\it third hint} and the firmest indication to date at ALP existence~\cite{gnrtb2023} (see also Section \ref{sec6.2} and~\cite{ALPinGRB2}). Moreover, all three above-mentioned hints are derived with the same ALP parameters, which further strengthens the indication itself.

In this review, we concentrate on the other main effect of the photon--ALP interaction, i.e., the change of the polarization state of photons. This ALP effect has attracted less interest so far with respect to that producing spectral modifications. However, the existing and proposed polarimetric missions in both the X-ray band such as IXPE~\cite{ixpe} (already operative), eXTP~\cite{extp}, XL-Calibur~\cite{xcalibur}, NGXP~\cite{ngxp}, XPP~\cite{xpp} and in the HE range like COSI~\cite{cosi} (approved to launch), e-ASTROGAM~\cite{eastrogam1,eastrogam2}, AMEGO~\cite{amego} allow polarization studies to be carried out with unprecedented sensitivity. Therefore, these observatories represent a unique opportunity to perform additional studies on ALP physics. In particular, the ALP effects on the polarization of photons produced in GRBs have been studied in~\cite{bassan}. Attention to various ALP-induced polarization topics regarding different astrophysical sources, such as AGN, white dwarfs, and neutron stars, has been paid in~\cite{ALPpol1,ALPpol2,ALPpolMWD,ALPpol3,ALPpol4,ALPpol5,day}. Moreover, the photon--ALP interaction has been demonstrated as a unique way to measure the {\it emitted} photon polarization~\cite{galantiTheorems}. Quite recently, the impact of the photon--ALP interaction on the polarization of photons produced in the central region of typical galaxy clusters or generated at the jet base of generic blazars has been analyzed in~\cite{galantiPol}. Results obtained in~\cite{galantiPol} have been applied to concrete cases: (i) to Perseus and Coma concerning galaxy clusters, as shown in~\cite{grtcClu}, and (ii) to OJ~287, BL~Lacertae, Markarian~501, and 1ES~0229+200 regarding blazars, as discussed in~\cite{grtPolBla}, showing that, in all the cases, the initial photon polarization is strongly modified by the photon--ALP interaction with the production of features observable by present and planned missions~\cite{ixpe,extp,xcalibur,ngxp,xpp,cosi,eastrogam1,eastrogam2,amego}. The above-cited studies concerning blazars~\cite{galantiPol,grtPolBla} and especially galaxy clusters~\cite{galantiPol,grtcClu} demonstrate that the measure of the polarization degree of photons produced by these astrophysical sources may provide us a new and an additional way to study ALPs and their effects. In fact, whenever the expectations of conventional physics about the photon polarization degree of astrophysical sources are in tension with possible future observations performed by missions in the X-ray band~\cite{ixpe,extp,xcalibur,ngxp,xpp} and in the HE range~\cite{cosi,eastrogam1,eastrogam2,amego}, this fact would represent a hint at ALP existence. This is especially true regarding galaxy clusters. As discussed below, photons diffusely produced in the central zone of galaxy clusters are expected to be unpolarized both in the X-ray and in the HE band within conventional physics, but the photon--ALP conversion makes them partially polarized, as shown in~\cite{galantiPol,grtcClu}. Therefore, a signal of partially polarized photons diffusely produced in a galaxy cluster would be a strong indication at ALP existence. Concerning blazars, we could achieve a similar conclusion if the detected polarization degree were extremely high, above the predictions of standard hadronic models (see below and~\cite{galantiPol,grtPolBla}). In case no tension was found, the ALP parameter space ($m_a$, $g_{a\gamma\gamma}$) could still be limited further. In this review, our aim is thus to describe the ALP-induced effects on photon polarization, which may be detected by missions in the X-ray band~\cite{ixpe,extp,xcalibur,ngxp,xpp} and in the HE range~\cite{cosi,eastrogam1,eastrogam2,amego} and which may provide us new ways to make progress in the ALP physics.

In particular, we introduce ALPs and the photon--ALP system in Section \ref{sec2}, and we illustrate polarization in Section \ref{sec3}. Section \ref{sec4} describes the astrophysical media crossed by the photon--ALP beam. In Section \ref{sec5}, the ALP-induced polarization effects are reported both in the X-ray and in the HE range for Perseus as an example for galaxy clusters and for OJ~287 as an example for blazars. In Section \ref{sec6}, the above-mentioned polarization effects are discussed along with their real detectability. Section \ref{sec6} is also devoted to describing possible future perspectives in the VHE range, where strong ALP signatures could be detected, along with a discussion about GRBs. Finally, we draw our conclusions in \mbox{Section \ref{sec7}}. {Throughout this review, we employ the natural Lorentz--Heaviside (rationalized) units with $\hslash = c = k_B = 1$.}

\section{Axion-like Particles}\label{sec2}

ALPs are neutral, spin-zero pseudo-scalar bosons, whose primary interaction with photons (other interactions with fermions or gluons may exist but are subdominant and negligible in the present context) is described by the {Lagrangian} 
\begin{eqnarray}
{\mathcal L}_{\rm ALP} =  \frac{1}{2} \, \partial^{\mu} a \, \partial_{\mu} a - \frac{1}{2} \, m_a^2 \, a^2 - \, \frac{1}{4 } g_{a\gamma\gamma} \, F_{\mu\nu} \tilde{F}^{\mu\nu} a  = \frac{1}{2} \, \partial^{\mu} a \, \partial_{\mu} a - \frac{1}{2} \, m_a^2 \, a^2 + g_{a\gamma\gamma} \, {\bf E} \cdot {\bf B}~a,
\label{lagr}
\end{eqnarray}
where $a$ denotes the ALP field, while ${\bf E}$ and ${\bf B}$ respectively represent the electric and magnetic components of the electromagnetic tensor $F_{\mu\nu}$, whose dual is $\tilde{F}^{\mu\nu}$.
Many bounds about the photon--ALP coupling $g_{a \gamma \gamma}$ and the ALP mass $m_a$ have been derived in the literature~\cite{cast,straniero,fermi2016,payez2015,berg,conlonLim,meyer2020,limFabian,limJulia,limKripp,limRey2,mwd}. The most reliable ALP limit is represented by $g_{a \gamma \gamma} < 0.66 \times 10^{- 10} \, {\rm GeV}^{- 1}$ for $m_a < 0.02 \, {\rm eV}$ at the $2 \sigma$ level from no detection of ALPs from the Sun~\cite{cast}. Other bounds are more or less affected by the description of the astrophysical environment~\cite{straniero,fermi2016,payez2015,berg,conlonLim,meyer2020,limFabian,limJulia,limKripp,limRey2,mwd}.

Furthermore, the QED one-loop vacuum polarization effects, which become important in the presence of a strong external magnetic field, are accounted by the Heisenberg--Euler--Weisskopf (HEW) effective Lagrangian~\cite{hew1, hew2, hew3}
\begin{equation}
\label{HEW}
{\mathcal L}_{\rm HEW} = \frac{2 \alpha^2}{45 m_e^4} \, \left[ \left({\bf E}^2 - {\bf B}^2 \right)^2 + 7 \left({\bf E} \cdot {\bf B} \right)^2 \right],
\end{equation}
where $\alpha$ represents the fine-structure constant, while $m_e$ is the electron mass.

We consider a photon--ALP beam of energy $E$ propagating in the $y$-direction in a magnetized medium with an external magnetic field denoted by ${\bf B}$ and entering \mbox{Equation~(\ref{lagr}}), where ${\bf E}$ describes the photon electric field. Because of the off-diagonality of the mass matrix of the $\gamma - a$ system, the propagation eigenstates differ from the interaction eigenstates, producing $\gamma \leftrightarrow a$ oscillations. This is similar to the oscillations of different flavor massive neutrinos, but in the case of the photon--ALP system, the external ${\bf B}$ field is essential to compensate for the spin mismatch between photons and ALPs. In particular, the functional expression of the photon--ALP coupling in Equation~(\ref{lagr}) shows that $a$ couples only with the component ${\bf B}_T$ of ${\bf B}$ transverse to the photon momentum $\bf k$~\cite{dgr2011}. Furthermore, the short-wavelength approximation~\cite{raffeltstodolsky} holds in the present analysis since $E \gg m_a$ (we consider extremely light ALPs), so that the propagation of an unpolarized photon--ALP beam arising from ${\mathcal L}_{\rm ALP}$ of Equation~(\ref{lagr}) is described by the Von Neumann-like equation
\begin{equation}
\label{vneum}
i \frac{d \rho (y)}{d y} = \rho (y) \, {\mathcal M}^{\dagger} ( E, y) - {\mathcal M} ( E, y) \, \rho (y),
\end{equation}
where ${\mathcal M} (E,y)$ represents the photon--ALP mixing matrix, while $\rho (y)$ is the polarization density matrix of the photon--ALP system. By defining $\phi$, the angle that ${\bf B}_T$ forms with the $z$ axis, $\mathcal M$ reads as follows:
\begin{eqnarray}
\label{mixmat}
{\mathcal M} (E,y) \equiv \left(
\begin{array}{ccc}
\Delta_{xx} (E,y) & \Delta_{xz} (E,y) & \Delta_{a \gamma}(y) \, {\rm sin} \, \phi \\
\Delta_{zx} (E,y) & \Delta_{zz} (E,y) & \Delta_{a \gamma}(y) \, {\rm cos} \, \phi \\
\Delta_{a \gamma}(y) \, {\rm sin}  \, \phi & \Delta_{ a \gamma}(y) \, {\rm cos} \, \phi & \Delta_{a a} (E) \\
\end{array}
\right),
\end{eqnarray}
with
\begin{equation}
\label{deltaxx}
\Delta_{xx} (E,y) \equiv \Delta_{\bot} (E,y) \, {\rm cos}^2 \, \phi + \Delta_{\parallel} (E,y) \, {\rm sin}^2 \, \phi,
\end{equation}
\begin{eqnarray}
\Delta_{xz} (E,y) = \Delta_{zx} (E,y) \equiv  \left(\Delta_{\parallel} (E,y) - \Delta_{\bot} (E,y) \right) {\rm sin} \, \phi \, {\rm cos} \, \phi,
\label{deltaxz}
\end{eqnarray}
\begin{equation}
\label{deltazz}
\Delta_{zz} (E,y) \equiv \Delta_{\bot} (E,y) \, {\rm sin}^2 \, \phi + \Delta_{\parallel} (E,y) \, {\rm cos}^2 \, \phi,
\end{equation}
\begin{equation}
\label{deltamix} 
\Delta_{a \gamma}(y) = \frac{1}{2}g_{a\gamma\gamma}B_T(y),
\end{equation}
\begin{equation}
\label{deltaM} 
\Delta_{aa} (E) = - \frac{m_a^2}{2 E},
\end{equation}
and
\begin{eqnarray}
\label{deltaort} 
\Delta_{\bot} (E,y) = \frac{i}{2 \, \lambda_{\gamma} (E,y)} - \frac{\omega^2_{\rm pl}(y)}{2 E}   + \frac{2 \alpha}{45 \pi} \left(\frac{B_T(y)}{B_{{\rm cr}}} \right)^2 E + \rho_{\rm CMB}E,
\end{eqnarray}
\begin{eqnarray}
\label{deltapar} 
\Delta_{\parallel} (E,y) = \frac{i}{2 \, \lambda_{\gamma} (E,y)} - \frac{\omega^2_{\rm pl}(y)}{2 E}  + \frac{7 \alpha}{90 \pi} \left(\frac{B_T(y)}{B_{{\rm cr}}} \right)^2 E + \rho_{\rm CMB}E ,    
\end{eqnarray}
\textls[-15]{where $B_{{\rm cr}} \simeq 4.41 \times 10^{13} \, {\rm G}$ is the critical magnetic field and $\rho_{\rm CMB}=0.522 \times 10^{-42}$. \mbox{Equation~(\ref{deltamix})}} describes the photon--ALP interaction, while Equation~(\ref{deltaM}) accounts for the ALP mass effect. The first term in Equations~(\ref{deltaort}) and~(\ref{deltapar}) describes photon absorption (e.g., due to the EBL~\cite{dgr2013,franceschinirodighiero,gptr}) with $\lambda_{\gamma}$ being the $\gamma\gamma \to e^+e^-$ mean free path. In the second term of Equations~(\ref{deltaort}) and~(\ref{deltapar}), $\omega_{\rm pl}=(4 \pi \alpha n_e / m_e)^{1/2}$ represents the plasma frequency with $n_e$ denoting the electron
number density. The third term in Equations~(\ref{deltaort}) and~(\ref{deltapar}) describes the QED one-loop vacuum polarization coming from ${\mathcal L}_{\rm HEW}$ of Equation~(\ref{HEW}), producing polarization variation and birefringence 
on the beam, and the fourth term expresses the photon dispersion on the CMB~\cite{raffelt2015}. 

The solution of Equation~(\ref{vneum}) can be expressed in terms of the {\it transfer matrix} ${\mathcal U}(E;y,y_0)$ of the photon--ALP system as
\begin{equation}
\label{unptrmatr}
\rho ( y ) = {\mathcal U} \bigl(E; y, y_0 \bigr) \, \rho_0 \, {\mathcal U}^{\dagger} \bigl(E; y, y_0 \bigr),
\end{equation}
where $\rho_0$ is the density matrix at position $y_0$. The probability to find the photon--ALP beam in the final state $\rho$ at position $y$ starting in
the initial state $\rho_0$ at position $y_0$ reads as follows:
\begin{equation}
\label{unpprob}
P_{\rho_0 \to \rho} (E,y) = {\rm Tr} \Bigl[\rho \, {\mathcal U} (E; y, y_0) \, \rho_0 \, {\mathcal U}^{\dagger} (E; y, y_0) \Bigr],
\end{equation}
with ${\rm Tr} \, \rho_0 = {\rm Tr} \, \rho =1$~\cite{dgr2011}.

We concentrate now on a photon--ALP system with the following characteristics: (i)~fully polarized photons, (ii) no photon absorption ($\lambda_{\gamma} \to \infty$), (iii) homogeneous medium, and (iv) constant $\bf B$ field. Therefore, we can choose the $z$ axis along the direction of ${\bf B}_T$, resulting in $\phi=0$ in Equation~(\ref{mixmat}). With these assumptions, the $\gamma \to a$ conversion probability reads as follows:
\begin{equation}
\label{convprob}
P_{\gamma \to a} (E, y) = \left(\frac{g_{a\gamma\gamma}B_T \, l_{\rm osc} (E)}{2\pi} \right)^2 {\rm sin}^2 \left(\frac{\pi (y-y_0)}{l_{\rm osc} (E)} \right),
\end{equation}
with
\begin{equation}
\label{losc}     
l_{\rm osc} (E) \equiv \frac{2 \pi}{\left[\bigl(\Delta_{zz} (E) - \Delta_{aa} (E) \bigr)^2 + 4 \, \Delta_{a\gamma}^2 \right]^{1/2}}~
\end{equation}
being the photon--ALP beam oscillation length. The different energy dependence of the several $\Delta$ terms in Equation~(\ref{mixmat}) produces various regimes, which can be experienced by the photon--ALP beam. Hence, we can introduce the {\it low-energy threshold} $E_L$ and the {\it high-energy threshold} $E_H$ reading as follows:
\begin{equation}
\label{EL}
E_L \equiv \frac{|m_a^2 - \omega^2_{\rm pl}|}{2 g_{a \gamma \gamma} \, B_T},  
\end{equation}
and 
\begin{equation}
\label{EH}
E_H \equiv g_{a \gamma \gamma} \, B_T \left[\frac{7 \alpha}{90 \pi} \left(\frac{B_T}{B_{\rm cr}} \right)^2 + \rho_{\rm CMB} \right]^{- 1},
\end{equation} 
respectively. In the energy interval $E_L \lesssim E \lesssim E_H$, the photon--ALP system lies in the {\it strong-mixing} regime, where the plasma, the ALP mass, the QED one-loop, and the photon dispersion on the CMB effects are all negligible with respect to the photon--ALP mixing term. In this case, $P_{\gamma \to a}$ is maximal and energy independent and reduces to
\begin{equation}
\label{convprobSM}
P_{\gamma \to a} (y) = {\rm sin}^2 \left( \frac{g_{a\gamma\gamma} B_T}{2}  (y-y_0) \right).
\end{equation}
Instead, both in the case $E \lesssim E_L$, where the plasma contribution and/or the ALP mass term are prominent, and in the case $E \gtrsim E_H$, where the QED one-loop effect and/or the photon dispersion on the CMB dominate, the photon--ALP beam propagates in the {\it weak-mixing} regime, where $P_{\gamma \to a}$ becomes energy dependent and progressively vanishes.

The usual situation of a generic photon initial polarization, structured magnetic fields, and non-homogeneous media represents a generalization of the above-considered case, which is reported here for the simplicity of the analytical expressions of the involved equations. However, in all our following calculations, we employ the appropriate photon polarization and the proper expressions of the magnetization, dispersion, and absorption properties of the crossed media.

\section{Polarization Effects}\label{sec3}

The generalized {\it polarization density matrix} $\rho$ associated with the photon--ALP beam can be expressed as 
\begin{equation}
\label{densmat}
\rho (y) = \left(\begin{array}{c}A_x (y) \\ A_z (y) \\ a (y)
\end{array}\right)
\otimes \left(\begin{array}{c}A_x (y) \  A_z (y) \ a (y) \end{array}\right)^{*},
\end{equation}
where $A_x (y)$ and $A_z (y)$ represent the photon linear polarization amplitudes along the $x$ and $z$ axis, respectively, and $a(y)$ is the ALP amplitude. Equation~(\ref{densmat}) allows us to describe totally polarized (pure states), unpolarized, and partially polarized beams at once. In particular, pure photon states polarized in the $x$ and $z$ direction are expressed as
\begin{equation}
\label{densphot}
{\rho}_x = \left(
\begin{array}{ccc}
1 & 0 & 0 \\
0 & 0 & 0 \\
0 & 0 & 0 \\
\end{array}
\right), \,\,\,\,\,\,\,\,
{\rho}_z = \left(
\begin{array}{ccc}
0 & 0 & 0 \\
0 & 1 & 0 \\
0 & 0 & 0 \\
\end{array}
\right),
\end{equation}
respectively, and the ALP state as
\begin{equation}
\label{densa}
{\rho}_a = \left(
\begin{array}{ccc}
0 & 0 & 0 \\
0 & 0 & 0 \\
0 & 0 & 1 \\
\end{array}
\right),
\end{equation}
whereas an unpolarized photon state is described by
\begin{equation}
\label{densunpol}
{\rho}_{\rm unpol} = \frac{1}{2} \left(
\begin{array}{ccc}
1 & 0 & 0 \\
0 & 1 & 0 \\
0 & 0 & 0 \\
\end{array}
\right).
\end{equation}
{Partially} 
 polarized photons are characterized by a polarization density matrix with an intermediate functional expression between Equations~(\ref{densphot}) and~(\ref{densunpol}). 

The photonic part of the polarization density matrix of Equation~(\ref{densmat}) can be expressed in terms of the Stokes parameters as~\cite{poltheor1}
\begin{equation}
\label{stokes}
{\rho}_{\gamma} = \frac{1}{2} \left(
\begin{array}{cc}
I+Q & U-iV \\
U+iV & I-Q \\
\end{array}
\right),
\end{equation}
while the photon degree of {\it linear polarization} $\Pi_L$ and the {\it polarization angle} $\chi$ are defined~\cite{poltheor2}~as
\begin{equation}
\label{PiL}
\Pi_L \equiv \frac{(Q^2+U^2)^{1/2}}{I} = \frac{\left[ (\rho_{11}-\rho_{22})^2+(\rho_{12}+\rho_{21})^2\right]^{1/2}}{\rho_{11}+\rho_{22}},
\end{equation}
and
\begin{equation}
\label{chiPol}
\chi \equiv \frac{1}{2}{\rm arctan}\left(\frac{U}{Q}\right)=\frac{1}{2}{\rm arctan}\left(\frac{\rho_{12}+\rho_{21}}{\rho_{11}-\rho_{22}}\right),
\end{equation}
respectively, where $\rho_{ij}$ with $i,j=1,2$ are the photon polarization density matrix elements. In the following Sections, we will show the behavior of the final $\Pi_L$, which turns out to be strongly modified by the photon--ALP interaction with respect to the initial photon degree of the linear polarization $\Pi_{L,0}$.

As shown by some theorems enunciated and demonstrated in~\cite{galantiTheorems}, in the absence of photon absorption, there exists a relationship between photon survival probability in the presence of photon--ALP interaction $P_{\gamma \to \gamma}$ and {\it initial} $\Pi_{L,0}$ ($P_{\gamma \to a}=1-P_{\gamma \to \gamma}$ due to the absence of photon absorption), and in particular, the following hold: 

\begin{enumerate}

\item In case of no photon absorption and a starting condition of only photons with an initial degree of linear polarization $\Pi_{L,0}$, the inequalities $P_{\gamma \to a} \le (1+\Pi_{L,0})/2$ and $P_{\gamma \to \gamma} \ge (1-\Pi_{L,0})/2$ hold. In case of initially unpolarized photons ($\Pi_{L,0}=0$), we observe $P_{\gamma \to a} \le 1/2$ and $P_{\gamma \to \gamma} \ge 1/2$.

\item In the previous conditions, $\Pi_{L,0}$ represents the measure of the overlap between the values assumed by $P_{\gamma \to a}$ and $P_{\gamma \to \gamma}$. In case of initially unpolarized photons ($\Pi_{L,0}=0$), $P_{\gamma \to a}$ and $P_{\gamma \to \gamma}$ possess the common value of 1/2, at most.

\end{enumerate}

We have employed the inequalities described in item 1 to verify the correctness of the behavior of $P_{\gamma \to \gamma}$ reported in several figures of this review (see the following sections).

As shown in~\cite{galantiTheorems}, the results presented in item 2 can be used to measure the {\it initial} $\Pi_{L,0}$ when only spectral data of an astrophysical source are known. The strategy works in the energy band in which the photon--ALP system lies in the weak mixing regime, where $P_{\gamma \to a}$ and $P_{\gamma \to \gamma}$ are energy dependent and their assumed values satisfy the inequalities described in item 1. 

The procedure can be sketched as follows. We consider systems in an energy range where photon absorption is negligible. Starting from observational data of an astrophysical source with the observed spectrum $\Phi_{\rm obs}$, we can infer $P_{\gamma \to \gamma}$ as 
\begin{equation}
\label{PsurvObsEm}
P_{\gamma \to \gamma}=\frac{\Phi_{\rm obs}}{\Phi_{\rm em}},
\end{equation}
where $\Phi_{\rm em}$ represents the emitted spectrum, which can be either known or derivable from $\Phi_{\rm obs}$ (see~\cite{galantiTheorems} for more details). Furthermore, the absence of photon absorption by hypothesis ensures that $P_{\gamma \to a} = 1 - P_{\gamma \to \gamma}$. Now, $\Pi_{L,0}$ can be obtained by measuring the interval of overlap between the values assumed by $P_{\gamma \to \gamma}$ and $P_{\gamma \to a}$.

In Figure~\ref{ALPsMeasurePol}, we show how the previous procedure works: in the left panels, an application in the optical-X-ray band is reported, while the right panels address a similar study but in the MeV range. We show how an observer would proceed to infer the initial $\Pi_{L,0}$ from spectral data. Note that cosmological effects are taken into account and $E$ in Figure~\ref{ALPsMeasurePol} represents the photon energy, as observed from the Earth.

In the optical-X-ray band, we consider a high-frequency peaked blazar (HBL) placed inside a poor galaxy cluster at redshift $z = 0.1$ with an initial $\Pi_{L,0}=0.3$, while in the MeV range, we deal with an HBL but now placed inside a quite rich galaxy cluster at redshift $z = 0.05$ with $\Pi_{L,0}=0.1$. We assume the phenomenological emitted spectra derived in~\cite{blazarSeq} and typical values concerning the magnetization properties of the astrophysical media crossed by the photon--ALP beam both in the optical-X-ray and in the MeV energy band.

{In particular, in the optical-X-ray band, the photon--ALP interaction turns out to be negligible inside the poor galaxy cluster (see also Section \ref{sec4.3}), where the blazar is supposed to be located and inside the hosting elliptical galaxy, whose typical properties are described in Section \ref{sec4.2}. Instead, we have efficient photon--ALP conversion inside the jet of the blazar, whose characteristics are described in Section \ref{sec4.1}: For the magnetic field in the jet ${\bf B}^{\rm jet}$, we assume a toroidal profile of Equation~(\ref{Bjet}) with magnetic field strength at the emission distance $y_{\rm em} = 3 \times 10^{16} \, \rm cm$ equal to $B_0^{\rm jet} \equiv B^{\rm jet}(y_{\rm em})= 0.5 \, \rm G$, and for the electron number density $n_e^{\rm jet}$, we consider the profile expressed by Equation~(\ref{njet}), taking $n_{e,0}^{\rm jet} \equiv n_e^{\rm jet}(y_{\rm em})=5 \times 10^4 \, \rm cm^{-3}$. We hypothesize an efficient photon--ALP interaction inside the extragalactic space by assuming an extragalactic magnetic field
strength $B_{\rm ext} = 1 \, \rm nG$ and the ${\bf B}_{\rm ext}$ coherence properties described in Section \ref{sec4.4}. Lastly, the photon--ALP interaction turns out to be efficient also inside the Milky Way, where we adopt the Jansson and Farrar model~\cite{jansonfarrar1,jansonfarrar2} concerning the magnetic field ${\bf B}_{\rm MW}$ and that presented in~\cite{yaomanchesterwang2017} regarding the electron number density in the Galaxy $n_{{\rm MW}, \, e}$ (see also Section \ref{sec4.5}). For more details, we send the reader to~\cite{galantiTheorems}.

\begin{figure}[H]
\hspace{-6pt}\includegraphics[width=0.5\textwidth]{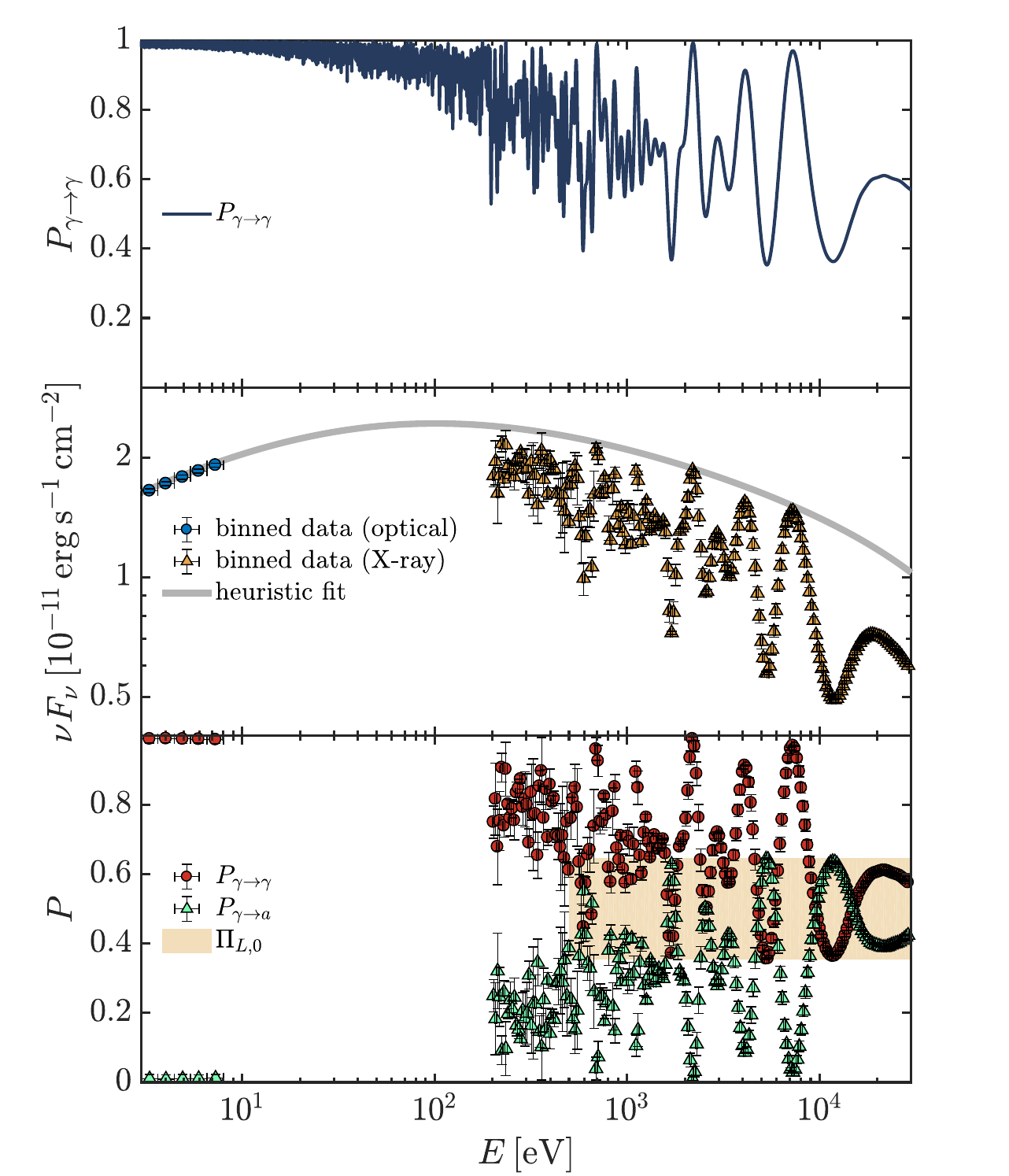}\includegraphics[width=0.5\textwidth]{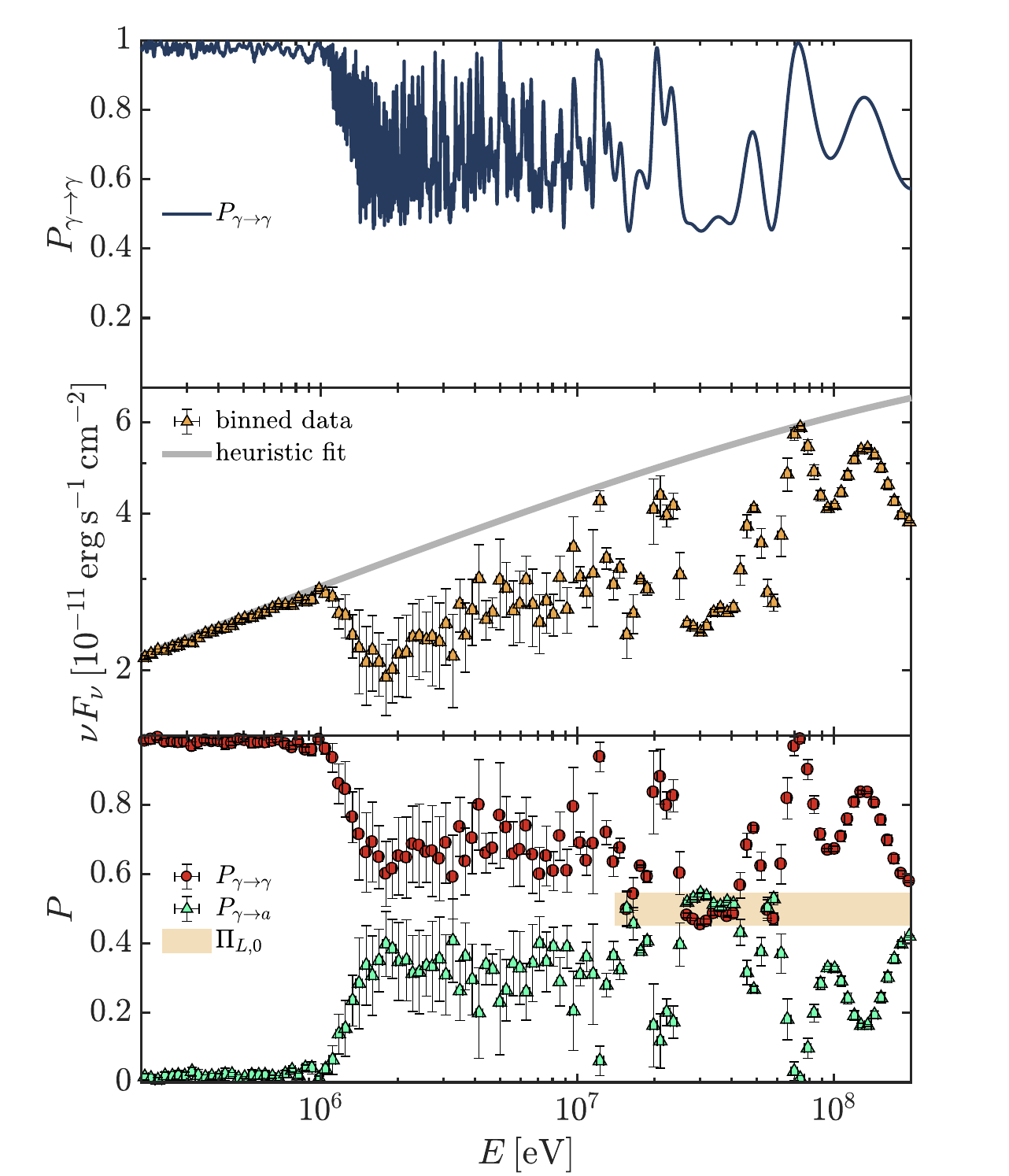}
\caption{ Measure of the initial $\Pi_{L,0}$. {\it Top panels}: A typical realization of $P_{\gamma \to \gamma}$ versus $E$. {\it Central panels}: Observed binned spectra. {\it Bottom panels}: Inferred $P_{\gamma \to \gamma}$ and $P_{\gamma \to a}$ and measure of $\Pi_{L,0}$. In the {\it left panels}, we consider a high-frequency peaked blazar (HBL) placed inside a poor galaxy cluster at redshift $z = 0.1$ with initial $\Pi_{L,0}=0.3$ in the energy range $3 \, {\rm eV} \le E \le 3 \times 10^4 \, {\rm eV}$, and we take $m_a = 5 \times 10^{-14} \rm\ eV$, $g_{a\gamma\gamma} = 0.5 \times 10^{-11} \, \rm GeV^{-1}$. In the {\it right panels}, we consider an HBL but now placed inside a quite rich galaxy cluster at redshift $z = 0.05$ with initial $\Pi_{L,0}=0.1$ in the energy range $2 \times 10^5 \, {\rm eV} \le E \le 2 \times 10^8 \, {\rm eV}$, and we take $m_a = 2 \times 10^{-10} \, \rm eV$, $g_{a\gamma\gamma} = 0.5 \times 10^{-11} \, \rm GeV^{-1}$. The inferred values of $\Pi_{L,0}$ are very close to the simulated ones. (Credit: adapted from~\cite{galantiTheorems}).}\label{ALPsMeasurePol}
\end{figure}

Instead, in the MeV range, we have an efficient photon--ALP conversion in the cluster since the blazar is supposed to be located inside a rich galaxy cluster, whose properties are reported in Section \ref{sec4.3}: we assume the cluster magnetic field ${\bf B}^{\rm clu}$ profile of Equation~(\ref{eq1}), where $n_e^{\rm clu}$ represents the cluster electron number density described by means of the $\beta$ model expressed by Equation~(\ref{eq2}). Concerning the parameters entering Equations~(\ref{eq1}) and~(\ref{eq2}), we assume $k_L = 0.1 \, \rm kpc^{-1}$, $k_H = 3 \, \rm kpc^{-1}$, $q = -11/3$, $B^{\rm clu}_0 = 20 \, \upmu{\rm G}$, $n_{e,0}^{\rm clu} = 0.1 \, \rm cm^{-3}$, $\eta_{\rm clu} = 0.75$, $\beta_{\rm clu} = 2/3$, a cluster core radius $r_{\rm core} = 150 \, \rm kpc$, and a cluster radius of $1 \, \rm Mpc$ (see Sections \ref{sec4.3} and \ref{sec6.1} for the definition of the parameters). In the present situation, we hypothesize a negligible photon--ALP conversion inside the extragalactic space by assuming $B_{\rm ext} < 10^{-15} \, \rm G$ (see also Section \ref{sec4.4}). Concerning the photon--ALP beam propagation inside the blazar jet, host galaxy, and Milky Way, we consider the same models and parameter values assumed in the previous case (optical-X-ray band) apart from the jet magnetic field strength at the emission distance, which is now $B_0^{\rm jet} \equiv B^{\rm jet}(y_{\rm em})= 0.1 \, \rm G$ (see also Sections~\ref{sec4.1}, \ref{sec4.2}, and \ref{sec4.5}). We send the reader to~\cite{galantiTheorems} for more details.}

In both of the previous cases (optical-X-ray and MeV band), a typical realization of $P_{\gamma \to \gamma}$ is evaluated and reported in the top panels of Figure~\ref{ALPsMeasurePol}: we assume $g_{a\gamma\gamma} = 0.5 \times 10^{-11} \, \rm GeV^{-1}$ for both of the situations, but we take $m_a = 5 \times 10^{-14} \, \rm eV$ for the first blazar in the optical-X-ray band, while we deal with an alternative scenario by taking $m_a = 2 \times 10^{-10} \, \rm eV$ for the second blazar in the MeV range. The central panels of Figure~\ref{ALPsMeasurePol} show the observed binned spectra $\Phi_{\rm obs}$ (with a typical energy resolution~\cite{optical,swift,eastrogam1,fermiSens}), simulated by multiplying the emitted spectra by $P_{\gamma \to \gamma}$ in the presence of photon--ALP interaction. In the bottom panels of Figure~\ref{ALPsMeasurePol}, we show $P_{\gamma \to \gamma}$ and $P_{\gamma \to a}$ with $P_{\gamma \to \gamma}$ inferred from the simulated observed spectra $\Phi_{\rm obs}$, reported in the central panels, by means of Equation~(\ref{PsurvObsEm}), where $\Phi_{\rm em}$ can be reconstructed by fitting the upper spectral bins in the central panels, since photon--ALP interaction produces an energy-dependent dimming in the observed spectra (see also~\cite{galantiTheorems}). We recall that $P_{\gamma \to a} = 1 - P_{\gamma \to \gamma}$ due to the absence of photon absorption. Finally, by employing the results presented in item 2, we can measure the initial $\Pi_{L,0}$, and we obtain $\Pi_{L,0} = 0.288 \pm 0.016$ in the X-ray band and $\Pi_{L,0} = 0.090 \pm 0.018$ in the MeV range. We observe that the inferred values are very close to the initially assumed ones, which demonstrates the power of the present method.

We want to stress that the above-described procedure is the only established possibility to measure the {\it initial} polarization of photons produced by astrophysical sources. Furthermore, this procedure can be employed in a wide energy range, where photon absorption is negligible, and only needs spectral observations of the source, so that flux-measuring observatories also become polarimeters.

\section{Photon--ALP Beam Propagation in Different Media}\label{sec4}

In the present section, we describe the main properties of the media crossed by the photon--ALP beam starting where photons are produced up to their arrival at the Earth: blazar, host galaxy, galaxy cluster, extragalactic space, and Milky Way. We concentrate mainly on the characteristics that are crucial for the photon--ALP system, such as the magnetization and absorption properties of the media, while we send the reader to the quoted papers for an exhaustive discussion.

\subsection{Blazars}\label{sec4.1}

\begin{figure}[H]
\includegraphics[width=0.6\textwidth]{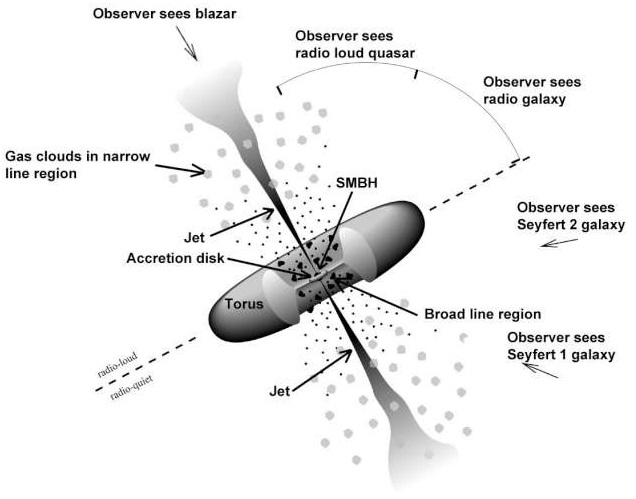}
\caption{Structure and phenomenology of active galactic nuclei. (Credit: adapted from~\cite{AGNfig1,AGNfig2,AGNfig3}).}\label{AGNphen}
\end{figure}

Blazars are powerful astrophysical sources emitting in the whole electromagnetic spectrum, from radio waves up to the VHE band. They are a particular class of AGN that are basically extragalactic supermassive black holes (SMBHs) accreting matter from the surroundings. In these systems, two collimated relativistic jets can develop in opposite directions. In the accidental case where one of these jets turns out to be  in the Earth direction, the AGN is called blazar, while in other situations, the AGN shows a different phenomenology, as displayed in Figure~\ref{AGNphen}, where the AGN structure is also shown. 

Blazars are divided into two sub-classes: flat spectrum radio quasars (FSRQs) and BL~Lac objects (BL~Lacs). FSRQs are powerful sources, where accretion from the disk surrounding the central SMBH is efficient. Furthermore, they show intense broad emission lines generated by clouds of gas in the so-called
broad-line region (BLR) encompassing the central SMBH and photoionized by the
ultraviolet photons produced by the disk. Moreover, FSRQs display absorption regions for VHE photons, which can interact with the optical/ultraviolet radiation emitted by the above-mentioned BLR and with the infrared radiation produced by the dusty torus (see Figure~\ref{AGNphen}), preventing the observation of photons with energy above $\sim$$20 \, \rm GeV$~\cite{FSRQabs1,FSRQabs2,FSRQabs3,gltcTorus}. However, photons of up to $\sim$$400 \, \rm GeV$ have been observed from FSRQs~\cite{FSRQvhe1,FSRQvhe2,FSRQvhe3}. Only ad hoc solutions have been proposed inside conventional physics, while photon--ALP oscillations provide a natural solution of the problem by partially preventing BLR absorption: this fact represents a hint at ALP existence~\cite{trgb2012}. Instead, BL Lacs are less powerful, and they show neither absorption regions nor significant emission lines due to the presence of a radiatively inefficient
accretion flow (RIAF) onto the central SMBH~\cite{riaf1,riaf2}.

In the following, we will concentrate on BL~Lacs, where the emission region of photons is located at a distance of about $y_{\rm em}= (10^{16} \text{--} 10^{17}) \, {\rm cm}$ from the central SMBH. Once produced at $y_{\rm em}$, photons propagate inside the BL Lac jet up to a distance $y_{\rm jet} \simeq 1 \, \rm kpc$, where they enter the host galaxy. During their path inside the jet, these photons can oscillate into ALPs inside the jet magnetic field ${\bf B}^{\rm jet}$. The toroidal part of ${\bf B}^{\rm jet}$, which is transverse to the jet axis~\cite{bbr1984,ghisellini2009,pudritz2011}, turns out to be the dominant one at the distances under consideration, and its profile reads as follows:
\begin{equation}
\label{Bjet}
B^{\rm jet} ( y ) = B^{\rm jet}_0 \left(\frac{y_{{\rm em}}}{y}\right),
\end{equation}
where $B^{\rm jet}_0$ denotes the magnetic field strength of the jet at $y_{\rm em}$. Another crucial information in order to evaluate photon--ALP conversion inside BL Lacs is the electron number density profile in the jet $n_e^{\rm jet}$, which, due to the conical shape of
the jet, is expected to be 
\begin{equation}
\label{njet}
n^{\rm jet}_e ( y ) = n^{\rm jet}_{e,0} \left(\frac{y_{{\rm em}}}{y}\right)^2,
\end{equation}
where $n^{\rm jet}_{e,0}$ represents the electron number density at $y_{\rm em}$. Synchrotron Self Compton (SSC) diagnostics applied to blazar spectra suggests assuming $n^{\rm jet}_{e,0}=5 \times 10^4 \, \rm cm^{-3}$~\cite{tavecchio2010}.

\begin{figure}[H]
\includegraphics[width=0.46\textwidth]{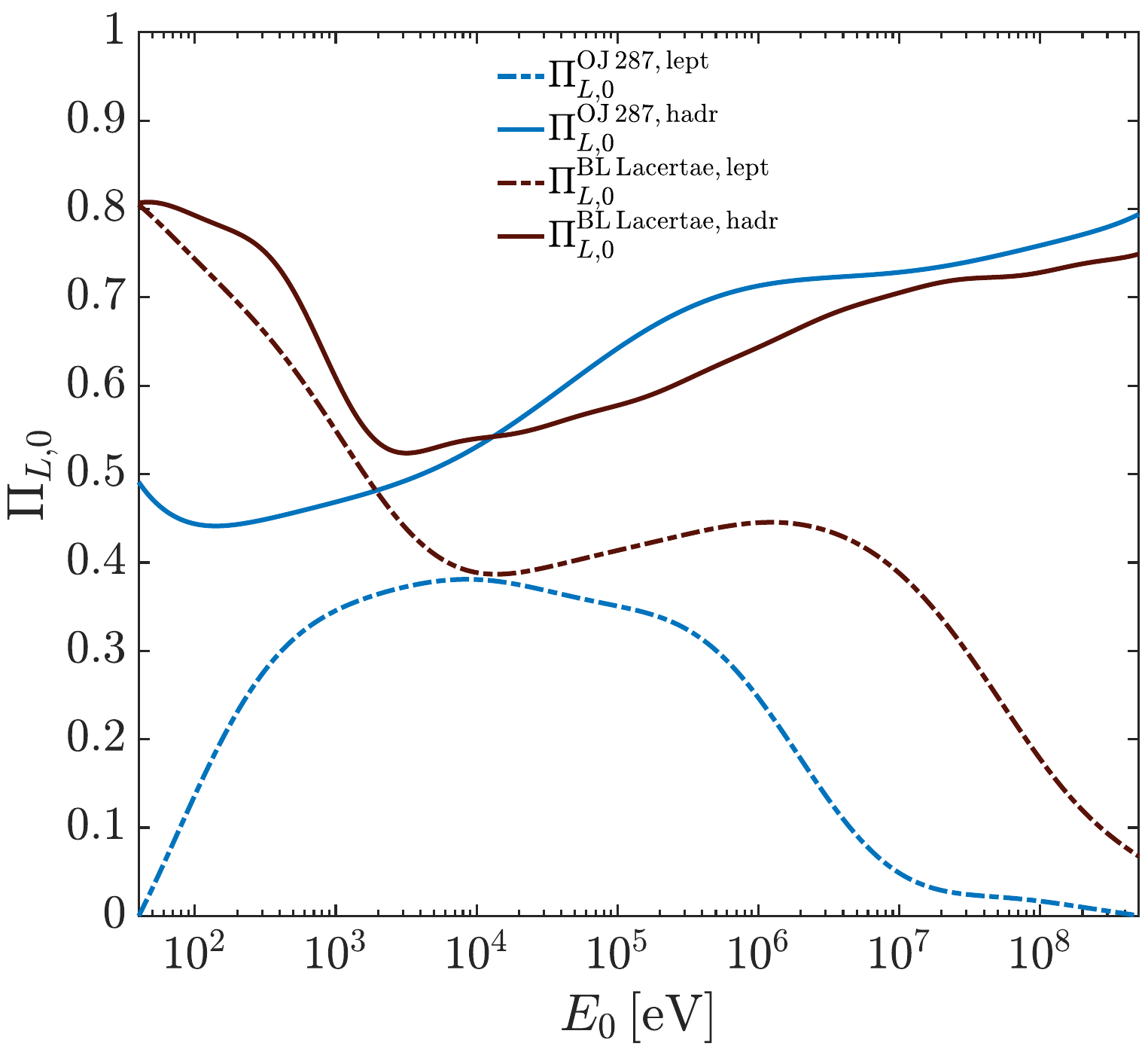}
\caption{\label{PiL0} Initial degree of linear polarization $\Pi_{L,0}$ for the blazars OJ~287 and BL~Lacertae in the case of both leptonic and hadronic models, as derived in~\cite{zhangBott}. (Credit:~\cite{grtPolBla}).}
\end{figure}

Two competing photon emission mechanisms have been proposed in order to explain blazar phenomenology: (i) the leptonic model (see, e.g.,~\cite{Maraschi92,Sikora94,ssc1}) and (ii) the hadronic model (see, e.g.,~\cite{mannheim1,mannheim2,Muecke2003}). Both scenarios describe photon production in the optical up to the X-ray band as generated by electron synchrotron emission. At higher energies, leptonic models explain photon emission with inverse Compton scattering of synchrotron radiation or of external photons from the disk and/or clouds with the synchrotron producing relativistic electrons~\cite{Maraschi92,Sikora94,ssc1}. Instead, hadronic models attribute the high-energy photon generation to proton--synchrotron emission or photomeson production~\cite{mannheim1,mannheim2,Muecke2003}. Hadronic models require higher values of $B^{\rm jet}_0$ and $y_{\rm em}$ with respect to leptonic scenarios. Moreover, hadronic models predict a higher initial degree of linear polarization $\Pi_{L,0}$ with respect to leptonic scenarios, as shown in Figure~\ref{PiL0} for two particular BL Lacs.

The propagation of the photon--ALP beam is computed in the comoving frame of the jet so that a transformation $E \to \gamma E$, with $\gamma$ being the Lorentz factor, is required to pass to the fixed frames of the subsequently encountered regions. 

The values assumed by $B^{\rm jet}_0$, $y_{\rm em}$, $\gamma$ and $\Pi_{L,0}$ affect the calculation of the photon--ALP beam propagation in the jet and will be discussed in Section \ref{sec5} for both leptonic and hadronic models. Then, the photon--ALP beam transfer matrix in the jet ${\mathcal U}_{\rm jet}$ can be evaluated by following the calculation reported in~\cite{trg2015}.

\subsection{Host Galaxy}\label{sec4.2}

The effectiveness of the photon--ALP conversion inside the host galaxy depends on its nature and characteristics. Ideal galaxies in order to have an efficient photon--ALP conversion are spirals and even more starbursts. The reason lies in the high magnetic field strength inside the host $B^{\rm host}$, which can reach values of $B^{\rm host}={\mathcal O}(5-10) \, \mu \rm G$ for spirals~\cite{SpiralBrev,Fletcher2010} and $B^{\rm host}={\mathcal O}(20-50) \, \mu \rm G$ for starbursts~\cite{Thompson2006,LopezRodriguez2021}, and in the large coherence length $L_{\rm dom}^{\rm host}$ of ${\bf B}^{\rm host}$, which is around $L_{\rm dom}^{\rm host} ={\mathcal O}(1-4) \, \rm kpc$ for both spiral and starburst galaxies outside the central region~\cite{SpiralBrev}. The magnetic field coherence can be modeled by employing a Kolmogorov-type turbulence power spectrum in a similar way as discussed in the following subsection concerning galaxy clusters (for more details about the parameter values, see also~\cite{gnrtb2023}). A possible example of photon--ALP conversion inside spirals or starbursts is represented by photons produced by GRBs, which are usually located in the central region of these galaxies~\cite{GRBposition,GRBposition2}. 

In particular, the location of GRB~221009A---the brightest GRB to date---in the central region of a disk-like galaxy~\cite{GRB221009Ahost} allows for an efficient photon--ALP conversion inside the host, which is able to justify the detection by the LHAASO Collaboration of this GRB above $10 \, \rm TeV$~\cite{LHAASO,LHAASOspectrumHigh,gnrtb2023}. The latter fact represents a strong hint at ALP existence, as demonstrated in~\cite{gnrtb2023} (see also Section \ref{sec6.2} and~\cite{ALPinGRB2}).

Instead, blazars are often hosted inside elliptical galaxies, where ${\bf B}^{\rm host}$ is described by means of a domain-like model with the typical values $B_{\rm host} \simeq 5 \, \upmu{\rm G}$ and $L_{\rm dom}^{\rm host} \simeq 150 \, {\rm pc}$~\cite{moss1996}. In such a situation, the photon--ALP oscillation length turns out to be much larger than $L_{\rm dom}^{\rm host}$, resulting in an inefficient photon--ALP conversion, as shown in~\cite{trgb2012}. However, whatever the nature and the properties of the host galaxy are, we evaluate the associated photon--ALP beam transfer matrix ${\mathcal U}_{\rm host}$.

\subsection{Galaxy Clusters}\label{sec4.3}

Galaxy clusters consist of 30 to more than 1000 galaxies with a total mass within the range $(10^{14}\text{--}10^{15}) \, M_{\odot}$, making them the largest gravitationally bound structures in the Universe. Properties such as galactic content, density profile, shape, concentration, and symmetry allow galaxy clusters to be classified into three main classes: (i) regular clusters, (ii) intermediate clusters, and (iii) irregular clusters~\cite{GalCluClass}. The approximate spherical symmetry and the satisfying knowledge of the density profile of regular clusters allow us to model their electron number density $n_e^{\rm clu}$ and the linked magnetic field ${\bf B}^{\rm clu}$ carefully so that we can properly calculate the photon--ALP conversion. 

The strength of ${\bf B}^{\rm clu}$ is set to $B^{\rm clu} = {\mathcal O} (1-10) \, \upmu{\rm G}$ by Faraday rotation measurements and synchrotron radio emission~\cite{cluB1,cluB2}. The morphology of ${\bf B}^{\rm clu}$ is turbulent and modeled with a Kolmogorov-type turbulence power spectrum $M(k)\propto k^q$, with index $q=-11/3$ and where $k$ represents the wave number in the interval $[k_L,k_H]$ with $k_L = 2\pi/\Lambda_{\rm max}$ and $k_H = 2\pi/\Lambda_{\rm min}$ being the minimal and maximal turbulence scales, respectively~\cite{cluFeretti}. Consequently, the behavior of ${\bf B}^{\rm clu}$ can be modeled as~\cite{cluFeretti,clu2}
\begin{equation}
\label{eq1}
B^{\rm clu}(y)={\mathcal B} \left( B_0^{\rm clu},k,q,y \right) \left( \frac{n_e^{\rm clu}(y)}{n_{e,0}^{\rm clu}} \right)^{\eta_{\rm clu}},
\end{equation}
where ${\mathcal B}$ represents the spectral function describing the Kolmogorov-type turbulence of the cluster magnetic field (for more details, see, e.g.,~\cite{meyerKolm}), $\eta_{\rm clu}$ is a cluster parameter, and $B_0^{\rm clu}$ and $n_{e,0}^{\rm clu}$ are the magnetic field strength and the electron number density at the center of the cluster, respectively.

Several models are employed to describe the profile of $n_e^{\rm clu}$, such as the single $\beta$ model typically used for non-cool-core (nCC) clusters, the double $\beta$ model for cool-core (CC) clusters, or a modified version of them~\cite{cluValues}. We will fix the behavior of $n_e^{\rm clu}$ in Section \ref{sec5}, where we will concentrate on real situations.

Photons can be produced by some astrophysical source such as a blazar or a GRB and can pass through a galaxy cluster, or they can be diffusely generated in the central region of the cluster itself. In the latter situation, it is necessary to discuss their initial $\Pi_{L,0}$. In the X-ray band, photons are produced in the cluster central region via thermal Bremsstrahlung~\cite{mitchell1979} with photons that turn out to be therefore unpolarized. In the HE range, several processes are proposed to explain possible photon emission, such as the cascade of VHE photons, inverse Compton scattering, and neutral pion decay (see, \mbox{e.g.,~\cite{cluGammaEm1,cluGammaEm2,cluGammaEm3,cluGammaEm4}}). Even in the HE range, photons turn out to be effectively \mbox{unpolarized~\cite{polarRev,FermiPol}}.

The transfer matrix ${\mathcal U}_{\rm clu}$ of the photon--ALP beam propagating from the central region of the cluster up to its virial radius can be evaluated by following the calculation reported in~\cite{galantiPol}. In the specific case in which photons are not diffusely produced in the central region of a rich cluster but are generated by another astrophysical source that does not lie inside a rich cluster---such a situation is frequent concerning blazars---the photon--ALP conversion in this region turns out to be inefficient.

\subsection{Extragalactic Space}\label{sec4.4}

The effectiveness of the photon--ALP conversion inside the extragalactic space is determined by the strength and morphology of the extragalactic magnetic field ${\bf B}_{\rm ext}$. However, the properties of ${\bf B}_{\rm ext}$ are still poorly understood. While many configurations for ${\bf B}_{\rm ext}$ exist in the literature~\cite{kronberg1994,grassorubinstein,wanglai}, its limits are in the interval $10^{- 7} \, {\rm nG} \lesssim B_{\rm ext} \lesssim 1.7 \, {\rm nG}$ on the scale of ${\mathcal O} (1) \, {\rm Mpc}$~\cite{neronovvovk,durrerneronov,upbbext}.

The shape of ${\bf B}_{\rm ext}$ is modeled as a domain-like network, where ${\bf B}_{\rm ext}$ is homogeneous over a whole domain of size $L_{\rm dom}^{\rm ext}$, which is equal to its coherence length, and with ${\bf B}_{\rm ext}$ randomly and {\it discontinuously} changing its direction from one domain to the subsequent one, but maintaining approximately the same strength~\cite{kronberg1994,grassorubinstein}. Since the exact orientation of ${\bf B}_{\rm ext}$ in each domain is unknown, the photon--ALP beam propagation becomes a {\it stochastic process}, and many realizations of it can be evaluated by varying the ${\bf B}_{\rm ext}$ orientations, in order to obtain the statistic properties of the system. However, only a single realization of the photon--ALP beam propagation process can be physically observed.

The domain-like model proposed for ${\bf B}_{\rm ext}$ is justified by several scenarios of outflows from primeval galaxies, further amplified by turbulence~\cite{reessetti1968,hoyle1969,kronbergleschhopp1999,furlanettoloeb2001}. According to the latter scenarios, the usual values for ${\bf B}_{\rm ext}$ are $B_{\rm ext} = {\mathcal O} (1) \, {\rm nG}$ and $L_{\rm dom}^{\rm ext} = {\mathcal O} (1) \, {\rm Mpc}$ (for details, see~\cite{grSM}). In this fashion, we assume $B_{\rm ext} = 1 \, {\rm nG}$ and $L_{\rm dom}^{\rm ext}$ in the interval $(0.2\text{--}10) \, \rm Mpc$ with $\langle L_{\rm dom}^{\rm ext}  \rangle = 2 \, \rm Mpc$.

However, the commonly used discontinuous model for ${\bf B}_{\rm ext}$ produces unphysical results when the photon--ALP beam oscillation length $l_{\rm osc}$ becomes smaller than $L_{\rm dom}^{\rm ext}$ due to the photon dispersion on the CMB~\cite{raffelt2015}, and consequently, the system turns out to be sensitive to the ${\bf B}_{\rm ext}$ substructure. In order to solve such a problem, we employ a more complex model developed in~\cite{grSM}, where the components of ${\bf B}_{\rm ext}$ change continuously, crossing from a domain to the subsequent one.

\textls[-15]{In the UV-X-ray and HE bands, photon absorption due to the interaction with EBL photons---we assume the conservative EBL model of Franceschini and Rodighiero~\cite{franceschinirodighiero}---is negligible, but it becomes substantial for $E > {\mathcal O}(100) \, \rm GeV$~\cite{dgr2013}. Therefore, for \mbox{$E < {\mathcal O}(100) \, \rm GeV$}}, propagation in the extragalactic space produces additional photon--ALP oscillations when the photon--ALP conversion is efficient. Instead, for $E > {\mathcal O}(100) \, \rm GeV$, where we have a large EBL absorption, the effect of the photon--ALP beam propagation in the extragalactic space is also to mitigate the EBL absorption of VHE photons. The reason lies in the fact that when photons become ALPs, they do not undergo EBL absorption so that the effective optical depth is reduced and more photons can reach the Earth.

In both of the previous situations, with large or negligible EBL absorption and in the case of both efficient or ineffective photon--ALP conversion, the transfer matrix ${\mathcal U}_{\rm ext}$ of the photon--ALP system in the extragalactic space can be calculated by following the analysis performed in~\cite{grExt,grSM}.

\subsection{Milky Way}\label{sec4.5}

Photon--ALP conversion in the Milky Way can be computed once the morphology and strength of the magnetic field ${\bf B}_{\rm MW}$ and the profile of the electron number density $n_{{\rm MW}, \, e}$ in the Galaxy are known.

Concerning ${\bf B}_{\rm MW}$, we employ the model developed by Jansson and Farrar~\cite{jansonfarrar1,jansonfarrar2}, which includes a disk and a halo component, both parallel to the Galactic plane, and a poloidal `X-shaped' component at the galactic center. The most relevant component of ${\bf B}_{\rm MW}$ for the photon--ALP conversion turns out to be its regular part, to which we add for completeness the turbulent one developed in~\cite{BMWturb}. The alternative ${\bf B}_{\rm MW}$ model of Pshirkov {et al.}~\cite{pshirkov2011} does not substantially produce a different photon--ALP conversion in the Milky Way. Still, we prefer the model by Jansson and Farrar~\cite{jansonfarrar1,jansonfarrar2} since it determines the Galactic halo component with a higher level of accuracy. 

Regarding $n_{{\rm MW}, \, e}$ we employ the model presented in~\cite{yaomanchesterwang2017}, where the Galaxy is modeled with an extended thick disk standing for the warm interstellar medium, a thin disk accounting for the Galactic molecular ring, a Galactic Center disk, spiral arms and features for the Gum Nebula, Galactic Loop I, and the Local Bubble.

By following the procedure developed in~\cite{gtre2019}, we can evaluate the transfer matrix ${\mathcal U}_{\rm MW}$ of the photon--ALP system in the Milky Way.

\subsection{Overall Photon--ALP Beam Propagation}

Once all the transfer matrices discussed in the previous subsections are known, we can combine them in order to calculate the total transfer matrix ${\mathcal U}$ describing the photon--ALP beam propagation starting from the photon emission zone up to the Earth. When photons are diffusely generated in the central zone of a galaxy cluster, ${\mathcal U}$ reads as follows:
\begin{equation} 
\label{Utot1}
{\mathcal U}={\mathcal U}_{\rm MW}\,{\mathcal U}_{\rm ext}\,{\mathcal U}_{\rm clu},
\end{equation}
while when photons are produced in the jet of a blazar, ${\mathcal U}$ is expressed by
\begin{equation} 
\label{Utot2}
{\mathcal U}={\mathcal U}_{\rm MW}\,{\mathcal U}_{\rm ext}\,{\mathcal U}_{\rm clu}\,{\mathcal U}_{\rm host}\,{\mathcal U}_{\rm jet}.
\end{equation}

In both of the previous cases, we can calculate the photon survival probability in the presence of photon--ALP interaction by specializing Equation~(\ref{unpprob}) as
\begin{equation} 
\label{probSurvFinal}
P_{\gamma \to \gamma}  = \sum_{i = x,z} {\rm Tr} \left[\rho_i \, {\mathcal U} \, \rho_{\rm in} \, {\mathcal U}^{\dagger} \right],
\end{equation}
where $\rho_x$ and $\rho_z$ read from Equations~(\ref{densphot}), and $\rho_{\rm in}$ represents the beam initial polarization density matrix, which is specified in the following subsections for concrete cases, for which the initial photon degree of linear polarization $\Pi_{L,0}$ is known. Moreover, by recalling Equation~(\ref{unptrmatr}) with $\rho_0 \equiv \rho_{\rm in}$, we obtain the final photon degree of the linear polarization $\Pi_L$ and the polarization angle $\chi$, thanks to Equations~(\ref{PiL}) and~(\ref{chiPol}), respectively (see also~\cite{galantiPol,grtcClu,grRew}).

\section{ALP Effects on Photon Polarization}\label{sec5}

Hereafter, we show how photon--ALP interaction modifies the final degree of linear polarization $\Pi_L$ and polarization angle $\chi$ for photons diffusely generated in the galaxy cluster central region or produced in the blazar jet. Here, we report two concrete examples: one for galaxy clusters, and specifically, we concentrate on Perseus, and one for BL Lacs and in particular, we focus on OJ~287. The latter two sources represent good observational targets to study ALP effects inside galaxy clusters~\cite{galantiPol,grtcClu} and blazars~\cite{galantiPol,grtPolBla}, respectively. We send the interested reader to the above-quoted papers, where a deeper discussion is presented along with other concrete examples of both galaxy clusters and blazars~\cite{galantiPol,grtcClu,grtPolBla}. We now describe the properties of Perseus and of OJ~287, which are crucial to evaluate the photon--ALP interaction effects on $\Pi_L$ and $\chi$.

Perseus is a deeply studied CC rich regular cluster located at redshift $z = 0.01756$ and represents the brightest galaxy cluster in the X-ray band. Its electron number density $n_e^{\rm clu}$ can be expressed by
\begin{eqnarray}
\label{eq2p}
&\displaystyle n_e^{\rm clu}(y)=n_{e,01}^{\rm clu} \left( 1+\frac{y^2}{r_{\rm core,1}^2} \right)^{-\frac{3}{2}\beta_{\rm clu,1}}+ 
n_{e,02}^{\rm clu} \left( 1+\frac{y^2}{r_{\rm core,2}^2} \right)^{-\frac{3}{2}\beta_{\rm clu,2}},
\end{eqnarray}
with $n_{e,01}^{\rm clu}=3.9 \times 10^{-2} \, \rm cm^{-3}$, $r_{\rm core,1}=80 \, \rm kpc$, $\beta_{\rm clu,1}=1.2$, $n_{e,02}^{\rm clu}=4.05 \times 10^{-3} \, \rm cm^{-3}$, $r_{\rm core,2}=280 \, \rm kpc$, and $\beta_{\rm clu,2}=0.58$~\cite{nePerseus}. The Perseus magnetic field ${\bf B}^{\rm clu}$ is described by Equation~(\ref{eq1}) by taking $B_0^{\rm clu} = 16 \, \upmu \rm G$, which represents an average estimate among those reported in the literature~\cite{BpersHigh,BpersLow}. Due to the similarity of the stochastic properties of the Perseus ${\bf B}^{\rm clu}$ with those of the Coma cluster, we assume the following values regarding Perseus: $q=-11/3$, $\Lambda_{\rm min}= 2 \, \rm kpc$ and $\Lambda_{\rm max}= 34 \, \rm kpc$~\cite{cluFeretti}, and the average value $\eta_{\rm clu}=0.5$ entering Equation~(\ref{eq1}). Finally, we take an initial degree of linear polarization $\Pi_{L,0}=0$ both in the UV-X-ray and in the HE band according to the discussion in Section~\ref{sec4.3}. Then, the photon--ALP beam propagation can be evaluated as discussed in Section \ref{sec4}, and its transfer matrix reads from Equation~(\ref{Utot1}).

OJ~287 represents a typical low-frequency peaked blazar (LBL) located at redshift $z=0.3056$. Due to its high flux in both the X-ray and HE bands, OJ~287 turns out to be a favorable observational target for polarimetric studies in both energy ranges~\cite{zhangBott}. In the following, we consider the typical values regarding the LBL parameters: for the leptonic model, we take $B^{\rm jet}_0 = 1 \, \rm G$, $y_{\rm em} = 3 \times 10^{16} \, \rm cm$ and $\gamma = 10$, while for the hadronic model, we assume $B^{\rm jet}_0 = 20 \, \rm G$, $y_{\rm em} = 10^{17} \, \rm cm$ and $\gamma = 15$~\cite{LeptHadrBott}. The profiles of ${\bf B}^{\rm jet}$ and $n_e^{\rm jet}$ are represented by Equations~(\ref{Bjet}) and~(\ref{njet}), respectively. As discussed in Section \ref{sec4.1}, the initial $\Pi_{L,0}$ for OJ~287 is plotted in Figure~\ref{PiL0} for both the leptonic and the hadronic scenario. OJ~287 is hosted inside an elliptical galaxy with a resulting inefficient photon--ALP conversion in the host, as discussed in Section \ref{sec4.2}. Moreover, OJ~287 is not found within a rich galaxy cluster with a consequent negligible ALP effect also in this zone (see \mbox{Section \ref{sec4.3}}). However, for completeness, we consider all the contributions for the evaluation of the total photon--ALP beam transfer matrix, which is expressed by Equation~(\ref{Utot2}). A remark regarding photons emitted by blazars should be kept in mind: due to the limited spatial resolution of real polarimeters, photons coming from different zones inside the transverse section of the blazar jet cannot be distinguished, and all photons are then collected together. Although the latter fact might in principle wash out the ALP-induced polarization features, the studies developed in~\cite{galantiPol,grtPolBla} demonstrate that this is not the case in realistic situations. In order to be conservative, the magnetic field profile expressed by Equation~(\ref{Bjet}) and a propagation distance of $1 \, \rm pc$ inside the jet are assumed. We send the reader to the papers quoted above~\cite{galantiPol,grtPolBla} for a deeper discussion regarding the latter problem and the adopted strategy to take the polarimeter technological limitations into account. 

We can now show the final photon survival probability $P_{\gamma \to \gamma}$, the corresponding photon degree of the linear polarization $\Pi_L$, and the polarization angle $\chi$ arising from photon--ALP oscillations inside the magnetized media analyzed in Section \ref{sec4} (blazar jet, host galaxy, galaxy cluster, extragalactic space, Milky Way) for the two above-discussed scenarios: (i)~photons are diffusely generated in the central zone of a galaxy cluster---we take Perseus as an example, and (ii) photons are produced at the jet base of a BL Lac---we consider OJ~287 as an illustration. In order to assess the robustness of the results, several realizations of the photon--ALP beam propagation process are evaluated by considering different magnetic field configurations due to limited knowledge especially regarding ${\bf B}^{\rm clu}$ and ${\bf B}_{\rm ext}$. The probability density function $f_{\Pi}$ associated with the final $\Pi_L$ can be computed by collecting all the values assumed by $\Pi_L$ in the various realizations and gives us information about the most probable scenarios.

Regarding the ALP parameters, we assume $g_{a\gamma\gamma}=0.5 \times 10^{-11} \, \rm GeV^{-1}$ and the two ALP masses: (i) $m_a \lesssim 10^{-14} \, \rm eV$ and (ii) $m_a = 10^{-10} \, \rm eV$, which are within the CAST bound~\cite{cast}. For $m_a \lesssim 10^{-14} \, \rm eV$, the plasma term dominates over the ALP mass term for the chosen model parameters, while the opposite is true in the case $m_a = 10^{-10} \, \rm eV$. Results about $P_{\gamma \to \gamma}$, $\Pi_L$ and $\chi$ are obtained for photons produced at redshift $z$ with emitted energy $E$ and observed energy $E_0=E/(1+z)$ in the two intervals: (i) UV-X-ray band and \mbox{(ii) HE band. }

Note that, in the following figures, binned data about $\Pi_L$ and $\chi$ are also shown in order to provide an idea of how a possible observation of ALP-induced features may look like. The binning procedure is based on the realistic energy resolution of current and/or planned polarimeters both in the X-ray band~\cite{ixpe,extp,xcalibur,ngxp,xpp} and in the HE range~\cite{cosi,eastrogam1,eastrogam2,amego} (see~\cite{grtcClu,grtPolBla} for more details).

\subsection{UV-X-ray Band}\label{sec5.1}

In the UV-X-ray band, the photon--ALP beam propagates in the weak mixing regime due to the strength of the plasma and/or ALP mass term inside the mixing matrix of Equation~(\ref{mixmat}). As a result, $P_{\gamma \to \gamma}$, $\Pi_L$ and $\chi$ turn out to possess an energy-dependent behavior in the case of both diffuse photon production in the central region of the Perseus cluster and photon emission at the jet base of OJ~287.

In particular, in the case $m_a = 10^{-10} \, \rm eV$, the ALP mass term is so strong that it dominates over the photon--ALP mixing term. As a result, the photon--ALP conversion turns out to be inefficient in the UV-X-ray band for $m_a = 10^{-10} \, \rm eV$, apart from a moderate dimming of $\Pi_{L,0}$ for OJ~287 in the case of the hadronic emission model due to the high central value of ${\bf B}^{\rm jet}$ (see~\cite{grtPolBla} for more details). This is the reason why we only consider the case $m_a \lesssim 10^{-14} \, \rm eV$ in this subsection. 

\begin{figure}[H]
\includegraphics[width=0.55\textwidth]{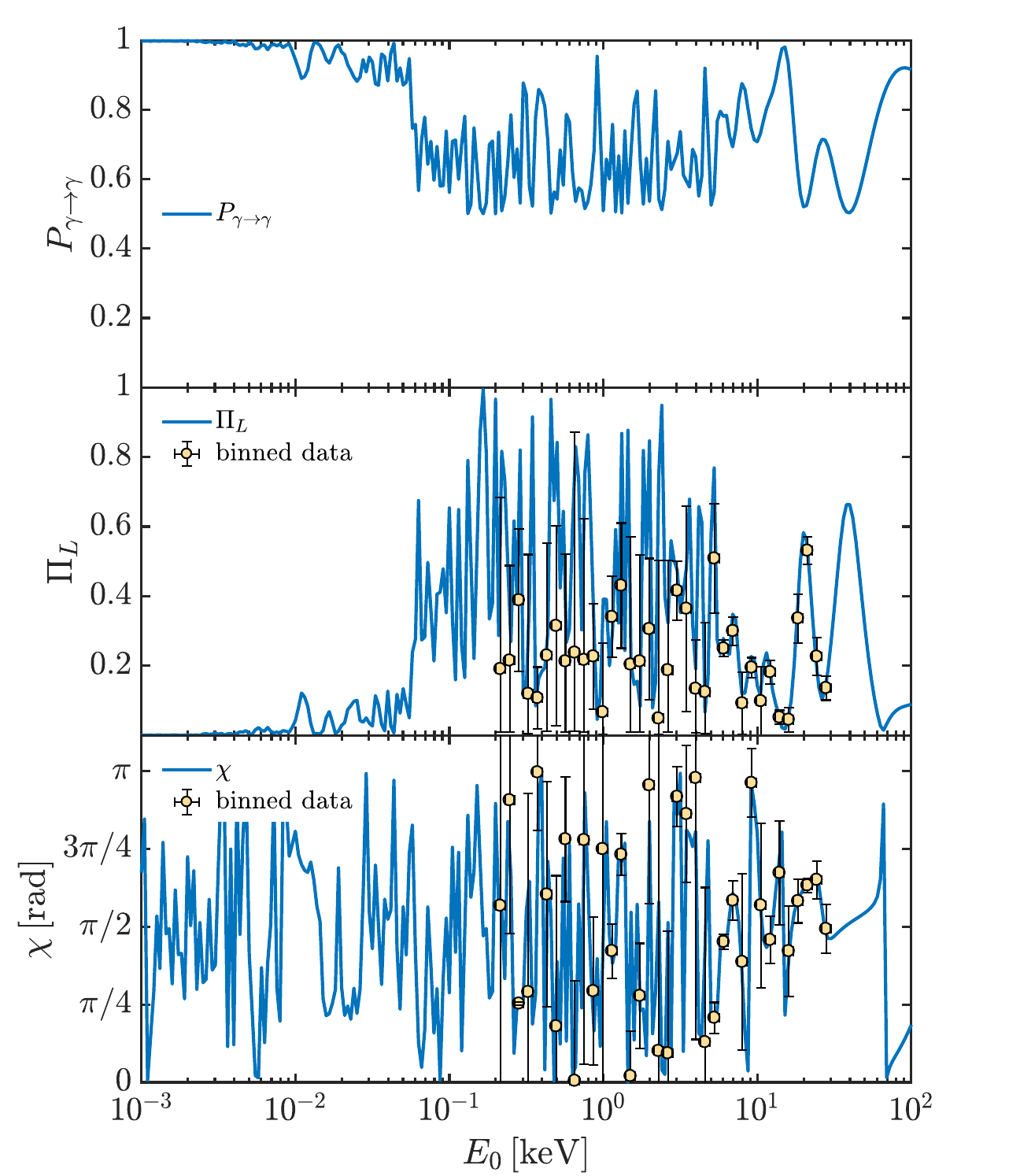}

\includegraphics[width=0.55\textwidth]{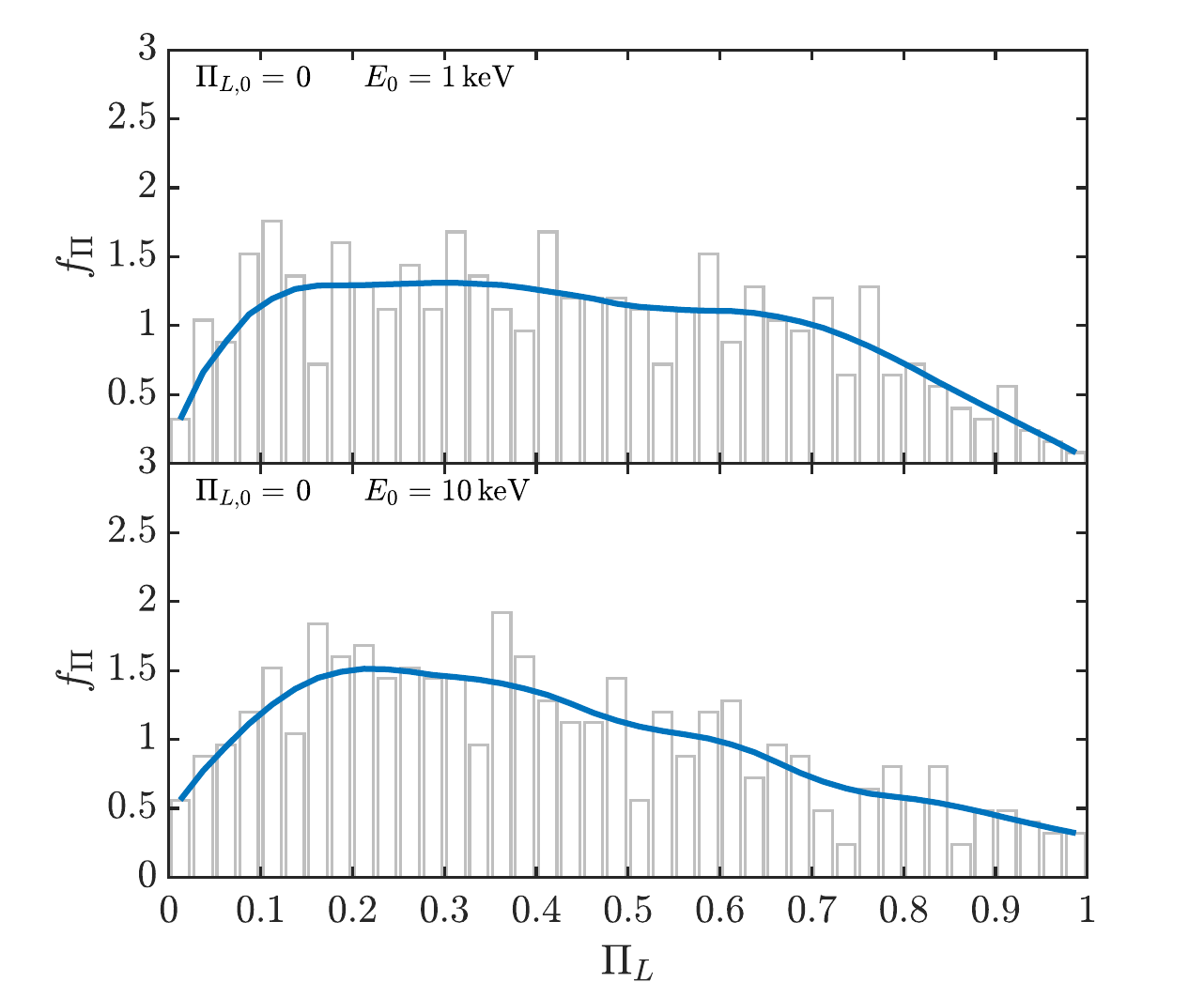}
\caption{Perseus cluster. ({\bf Top subfigure}) Photon survival probability $P_{\gamma \to \gamma }$ (upper panel), corresponding final degree of linear polarization $\Pi_L$ (central panel), and final polarization angle $\chi$ (lower panel) in the energy range $(10^{-3}\text{--}10^2) \, {\rm keV}$. We take $g_{a\gamma\gamma}=0.5 \times 10^{-11} \, \rm GeV^{-1}$, $m_a \lesssim 10^{-14} \, \rm eV$. The initial degree of linear polarization is $\Pi_{L,0}=0$. ({\bf Bottom subfigure}) Probability density function $f_{\Pi}$ obtained by interpolating the plotted histogram for several realizations of $\Pi_L$ at $1 \, \rm keV$ (upper panel) and $10 \, \rm keV$ (lower panel). (Credit:~\cite{grtcClu}).}\label{perseusX-14}
\end{figure}

\begin{figure}[H]
\includegraphics[width=0.956\textwidth]{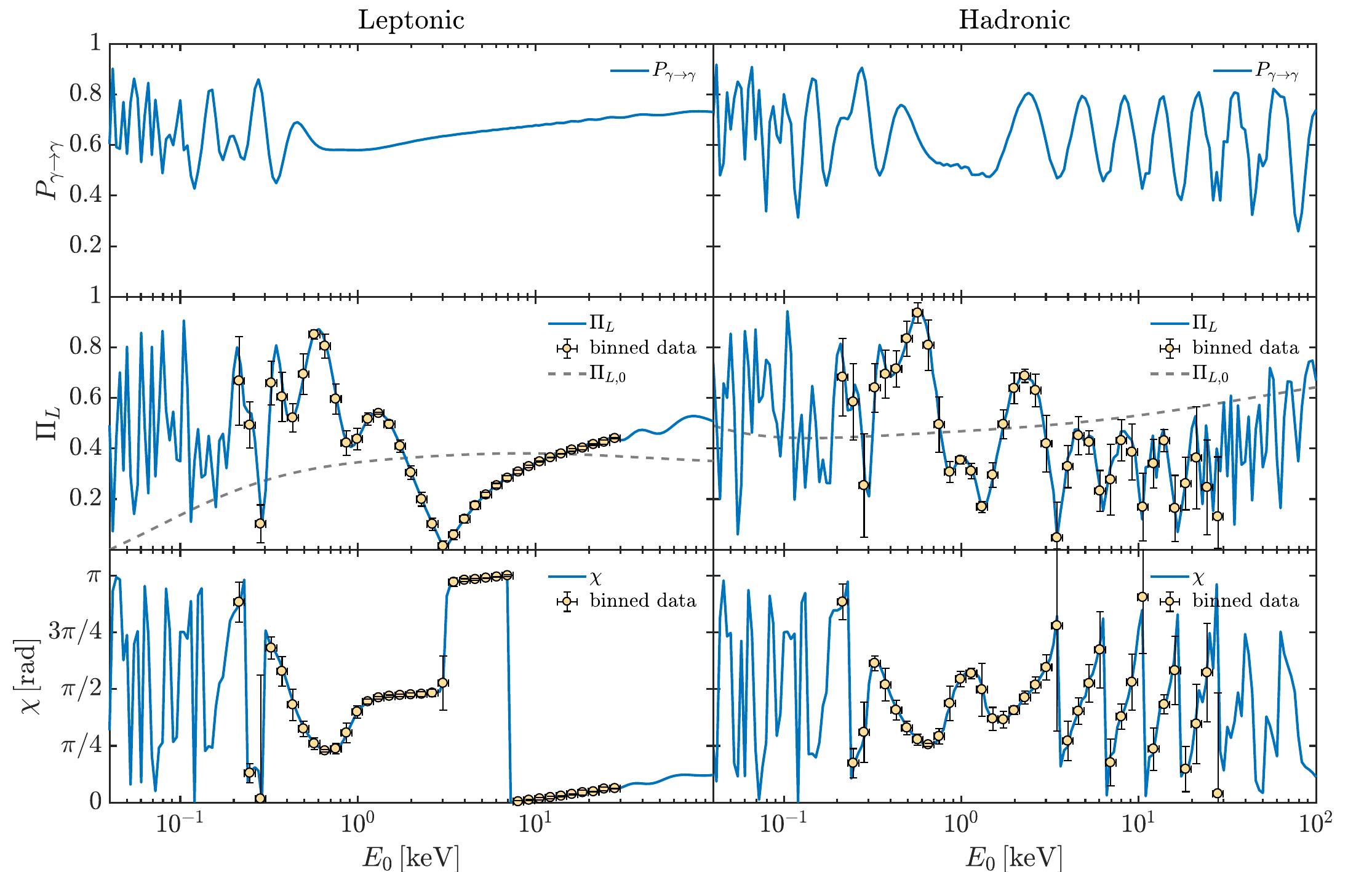}

\includegraphics[width=0.956\textwidth]{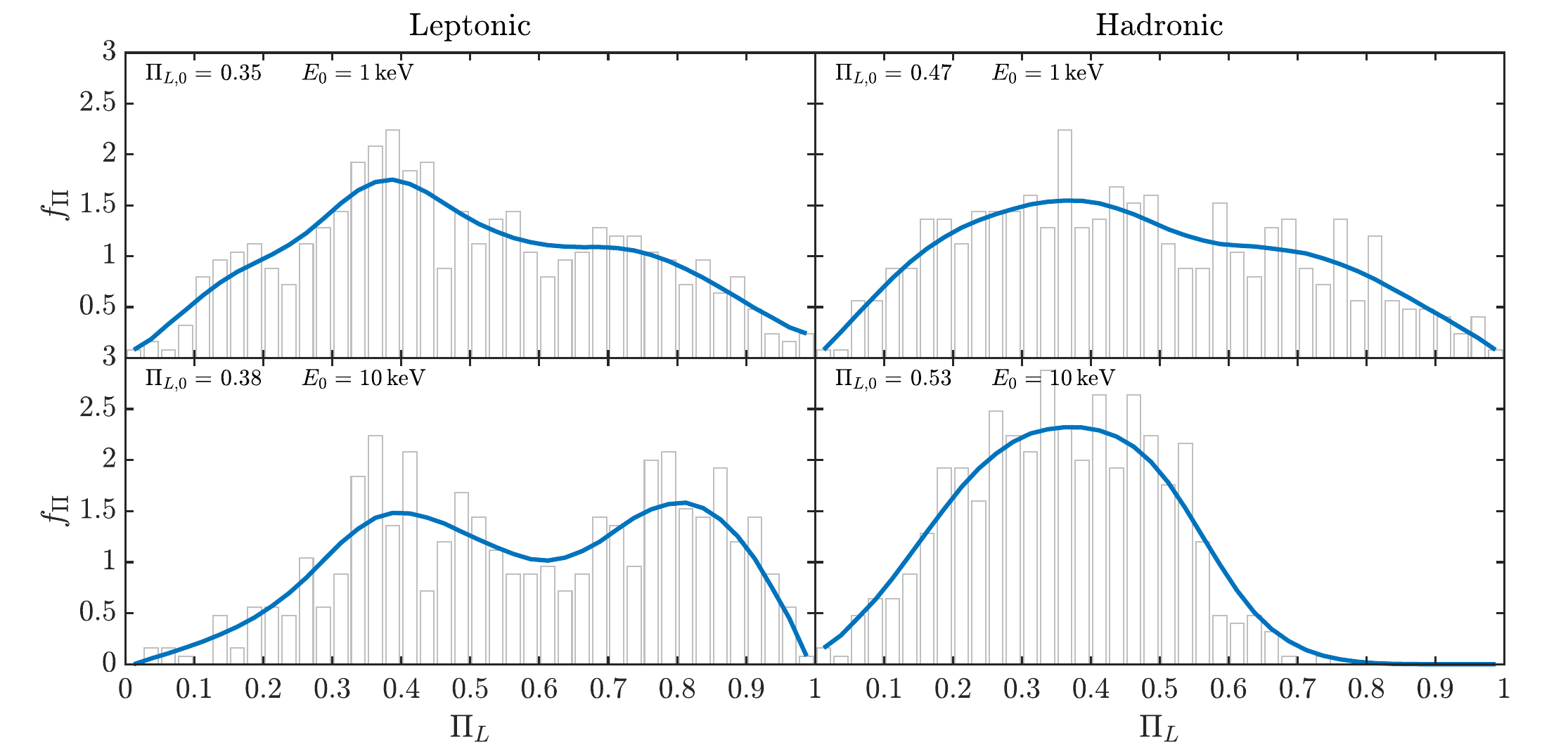}
\caption{OJ~287. We consider a leptonic and a hadronic emission mechanism in the left and right columns, respectively. Correspondingly, the initial degree of linear polarization $\Pi_{L,0}$ is also shown (see also Figure~\ref{PiL0}). ({\bf Top subfigure}) Photon survival probability $P_{\gamma \to \gamma }$ (upper panels), corresponding final degree of linear polarization $\Pi_L$ (central panels), and final polarization angle $\chi$ (lower panels) in the energy range $(4 \times 10^{-2}\text{--}10^2) \, {\rm keV}$. We take $g_{a\gamma\gamma}=0.5 \times 10^{-11} \, \rm GeV^{-1}$, $m_a \lesssim 10^{-14} \, \rm eV$. ({\bf Bottom subfigure}) Probability density function $f_{\Pi}$ obtained by interpolating the plotted histogram for several realizations of $\Pi_L$ at $1 \, \rm keV$ (upper panels) and $10 \, \rm keV$ (lower panels). (Credit:~\cite{grtPolBla}).}\label{OJ287X-14}
\end{figure}

Regarding the Perseus cluster, we report $P_{\gamma \to \gamma}$, $\Pi_L$ and $\chi$ in the top subfigure of Figure~\ref{perseusX-14} and the corresponding $f_{\Pi}$ associated with the final $\Pi_L$ in the bottom subfigure (see~\cite{grtcClu} for more details). Similarly, concerning OJ~287, we show $P_{\gamma \to \gamma}$, $\Pi_L$ and $\chi$ in the top subfigure of Figure~\ref{OJ287X-14} and the linked $f_{\Pi}$ in the bottom subfigure, but we now consider the two possible emission mechanisms inside blazars (see Section \ref{sec4.1}): (i) the leptonic model in the left panels and (ii) the hadronic model in the right panels (see~\cite{grtPolBla} for more details). In both cases, i.e., for both Perseus and OJ~287, we concentrate on two benchmark energies in the evaluation of $f_{\Pi}$: (i) $E_0 = 1 \, \rm keV$ and (ii) $E_0 = 10 \, \rm keV$.

In the case considered here, i.e., $g_{a\gamma\gamma}=0.5 \times 10^{-11} \, \rm GeV^{-1}$, $m_a \lesssim 10^{-14} \, \rm eV$, the energy dependence of $P_{\gamma \to \gamma}$, $\Pi_L$, and $\chi$ in the top subfigures of Figures~\ref{perseusX-14} and~\ref{OJ287X-14} is caused by the strength of the plasma term, which is not negligible with respect to the photon--ALP mixing term, especially inside the cluster in the case of Perseus and in the BL Lac jet regarding OJ~287. As a result, the photon--ALP beam propagates in the weak mixing regime, which protracts for a few energy decades due to the large variation with respect to the distance of ${\bf B}^{\rm clu}$ and $n_e^{\rm clu}$ regarding Perseus and of ${\bf B}^{\rm jet}$ and $n_e^{\rm jet}$ concerning OJ~287 (see \mbox{Sections \ref{sec4.1} and \ref{sec4.3}} and~\cite{galantiPol,grtcClu,grtPolBla} for details). In addition, the previous figures demonstrate that the final $\Pi_L$ turns out to be heavily modified with respect to the initial one $\Pi_{L,0}$ for both Perseus and OJ~287. The binned data in the top Subfigures of Figures~\ref{perseusX-14} and~\ref{OJ287X-14} concerning $\Pi_L$ and $\chi$ show that the ALP-induced features can be detected by IXPE~\cite{ixpe}, eXTP~\cite{extp}, XL-Calibur~\cite{xcalibur}, NGXP~\cite{ngxp}, and XPP~\cite{xpp} for $E_0 \gtrsim 2 \, \rm keV$ regarding Perseus and for $E_0 \gtrsim 0.5 \, \rm keV$ concerning OJ~287. Regarding OJ~287, the top subfigure of Figure~\ref{OJ287X-14} shows that the hadronic model produces a larger energy variability of $\Pi_L$ and $\chi$ with respect to the leptonic scenario due to the higher value of ${B}^{\rm jet}_0$ and the wider extension of the jet~region.

The behavior of $f_{\Pi}$ reported in the bottom subfigures of Figures~\ref{perseusX-14} and~\ref{OJ287X-14} shows a broadening of the final $\Pi_L$ with respect to the initial $\Pi_{L,0}$ for both Perseus and OJ~287. Perseus appears as a strong target to study ALP effects on photon polarization, since the final value $\Pi_L = 0$ is never the most probable expected result, while conventional physics predicts $\Pi_L = \Pi_{L,0} = 0$. Instead, for OJ~287, the most probable final value of $\Pi_L$ is not very different from the initial $\Pi_{L,0}$ apart from the case of the leptonic model at $E_0 = 10 \, \rm keV$, where the most probable result turns out to be $\Pi_L \gtrsim 0.8$, which is extremely difficult to explain within conventional physics. For more details, we send the reader to the discussion developed in~\cite{grtcClu} concerning Perseus and to that presented in~\cite{grtPolBla} regarding OJ~287.

\subsection{High-Energy Band}\label{sec5.2}

In the HE band, both the scenario with ALP parameters $g_{a\gamma\gamma}=0.5 \times 10^{-11} \, \rm GeV^{-1}$, $m_a \lesssim 10^{-14} \, \rm eV$ and that with $g_{a\gamma\gamma}=0.5 \times 10^{-11} \, \rm GeV^{-1}$, $m_a = 10^{-10} \, \rm eV$ produce sizable ALP effects on $P_{\gamma \to \gamma}$, $\Pi_L$ and $\chi$ in the case both of photon diffuse production in the central zone of the Perseus cluster and of photon emission at the jet base of OJ~287.

\begin{figure}[H]
\includegraphics[width=0.55\textwidth]{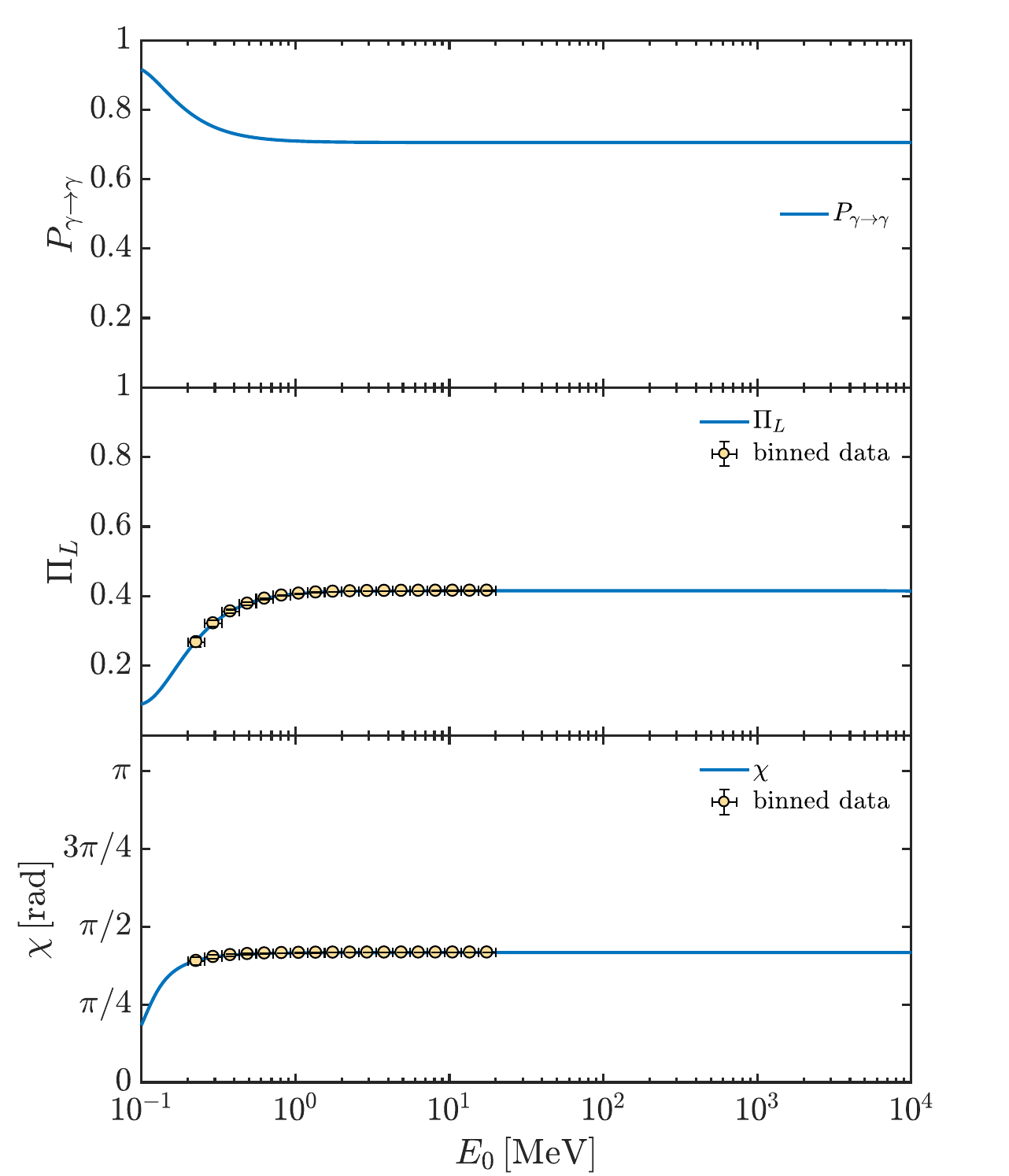}

\includegraphics[width=0.55\textwidth]{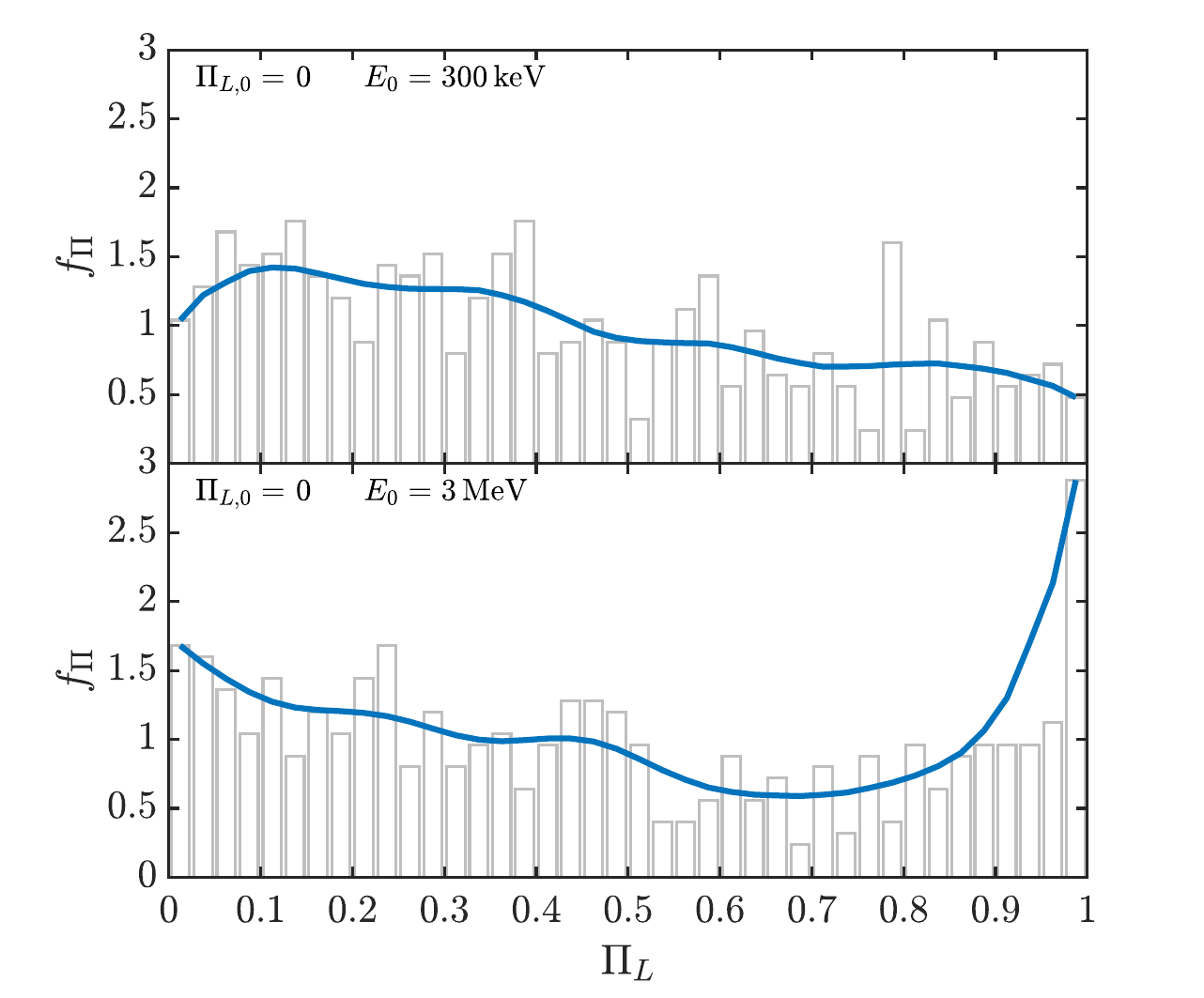}
\caption{Perseus cluster. ({\bf Top subfigure}) Photon survival probability $P_{\gamma \to \gamma }$ (upper panel), corresponding final degree of linear polarization $\Pi_L$ (central panel), and final polarization angle $\chi$ (lower panel) in the energy range $(10^{-1}\text{--}10^4) \, {\rm MeV}$. We take $g_{a\gamma\gamma}=0.5 \times 10^{-11} \, \rm GeV^{-1}$, $m_a \lesssim 10^{-14} \, \rm eV$. The initial degree of linear polarization is $\Pi_{L,0}=0$. ({\bf Bottom subfigure}) Probability density function $f_{\Pi}$ obtained by interpolating the plotted histogram for several realizations of $\Pi_L$ at $300 \, \rm keV$ (upper panel) and $3 \, \rm MeV$ (lower panel). (Credit:~\cite{grtcClu}).}\label{perseusHE-14}
\end{figure}

\begin{figure}[H]
\includegraphics[width=0.55\textwidth]{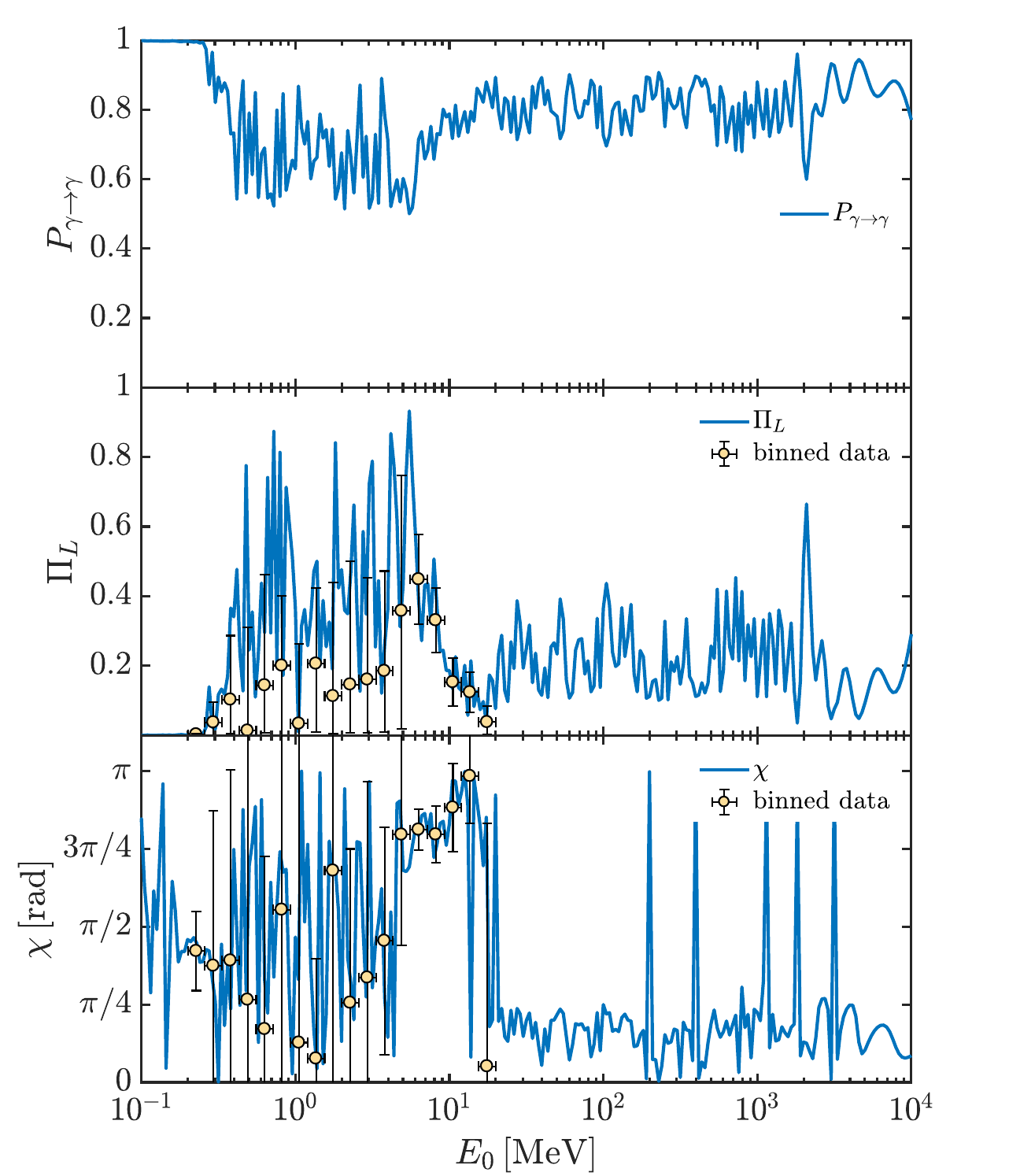}

\includegraphics[width=0.55\textwidth]{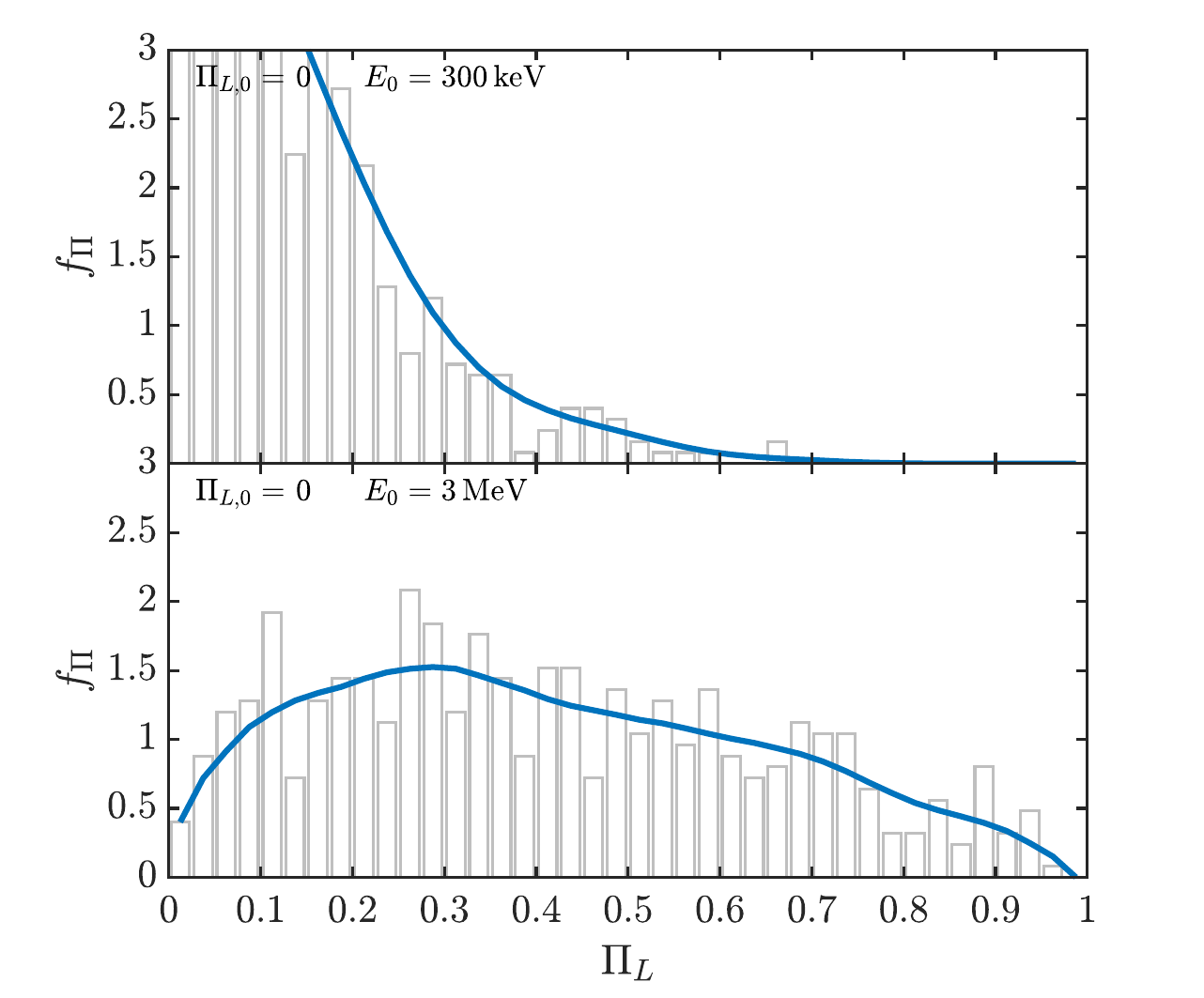}
\caption{Perseus cluster. Same as Figure~\ref{perseusHE-14}. We take $g_{a\gamma\gamma}=0.5 \times 10^{-11} \, \rm GeV^{-1}$, $m_a = 10^{-10} \, \rm eV$. The initial degree of linear polarization is $\Pi_{L,0}=0$. (Credit:~\cite{grtcClu}).}\label{perseusHE-10}
\end{figure}

\begin{figure}[H]
\includegraphics[width=0.956\textwidth]{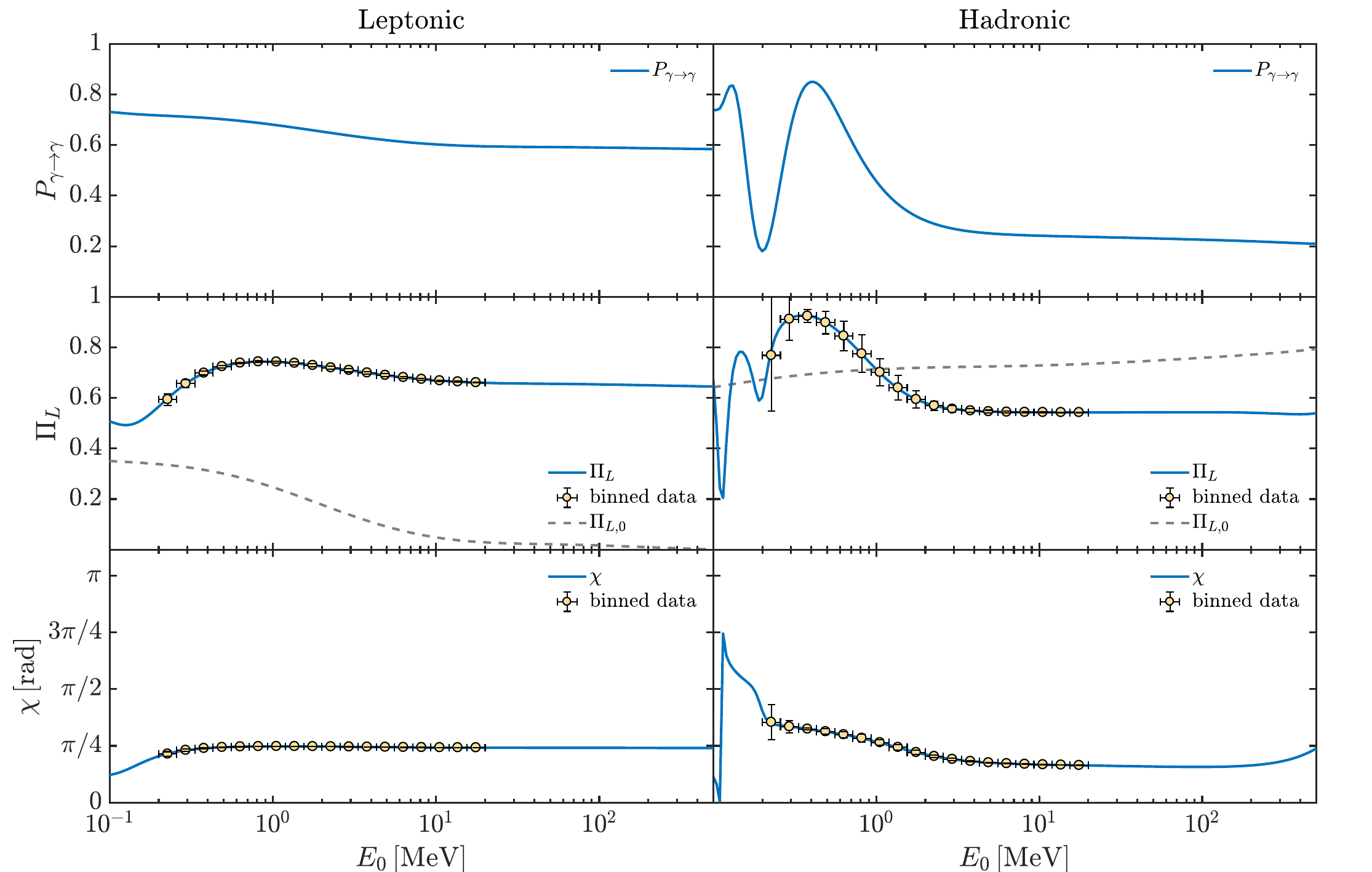}

\includegraphics[width=0.956\textwidth]{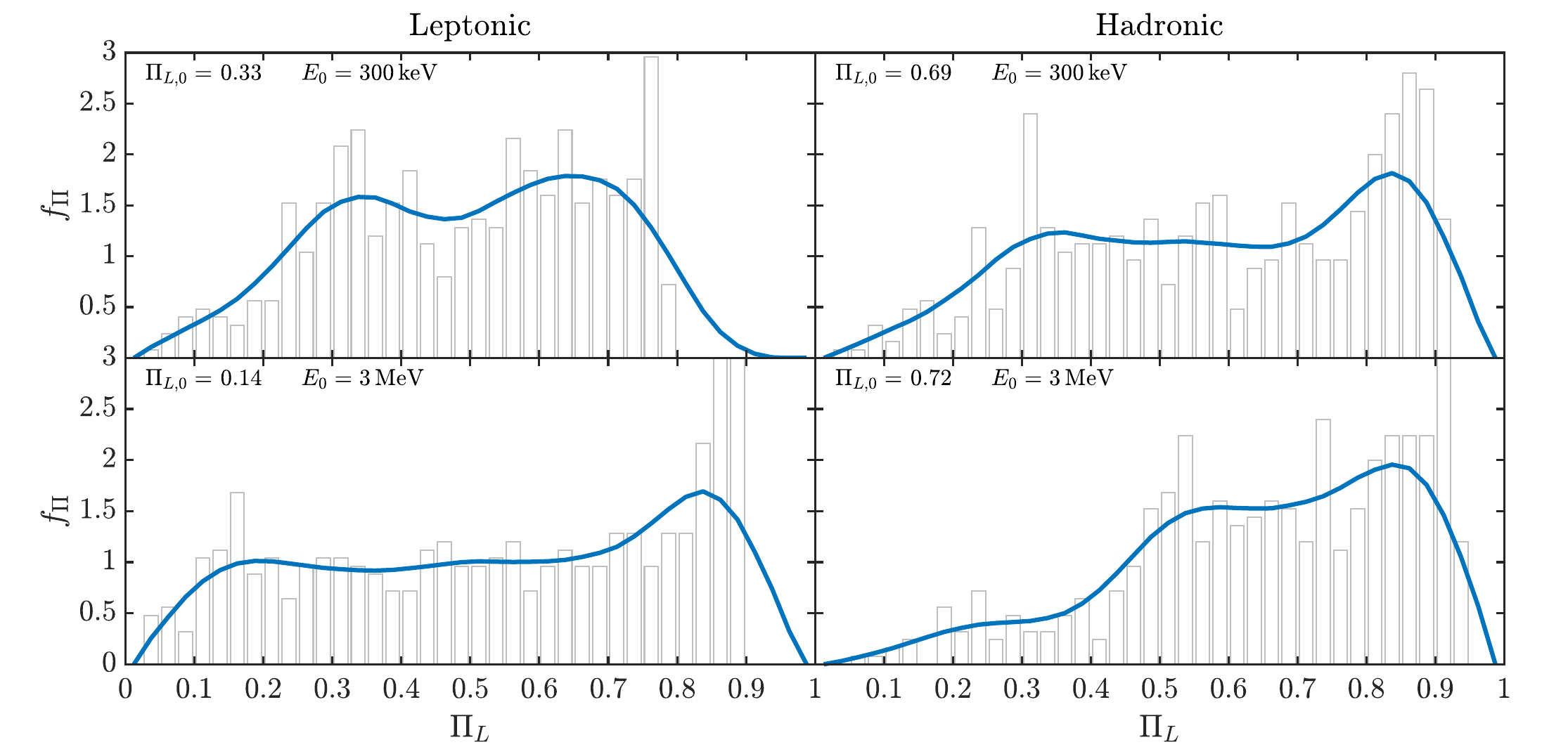}
\caption{OJ~287. We consider a leptonic and a hadronic emission mechanism in the left and right columns, respectively. Correspondingly, the initial degree of linear polarization $\Pi_{L,0}$ is also shown (see also Figure~\ref{PiL0}). ({\bf Top subfigure}) Photon survival probability $P_{\gamma \to \gamma }$ (upper panels), corresponding final degree of linear polarization $\Pi_L$ (central panels), and final polarization angle $\chi$ (lower panels) in the energy range $(10^{-1} \text{--}5 \times 10^2) \, {\rm MeV}$. We take $g_{a\gamma\gamma}=0.5 \times 10^{-11} \, \rm GeV^{-1}$, $m_a \lesssim 10^{-14} \, \rm eV$. ({\bf Bottom subfigure}) Probability density function $f_{\Pi}$ obtained by interpolating the plotted histogram for several realizations of $\Pi_L$ at $300 \, \rm keV$ (upper panels) and $3 \, \rm MeV$ (lower panels). (Credit:~\cite{grtPolBla}).}\label{OJ287HE-14}
\end{figure}

\begin{figure}[H]
\includegraphics[width=0.956\textwidth]{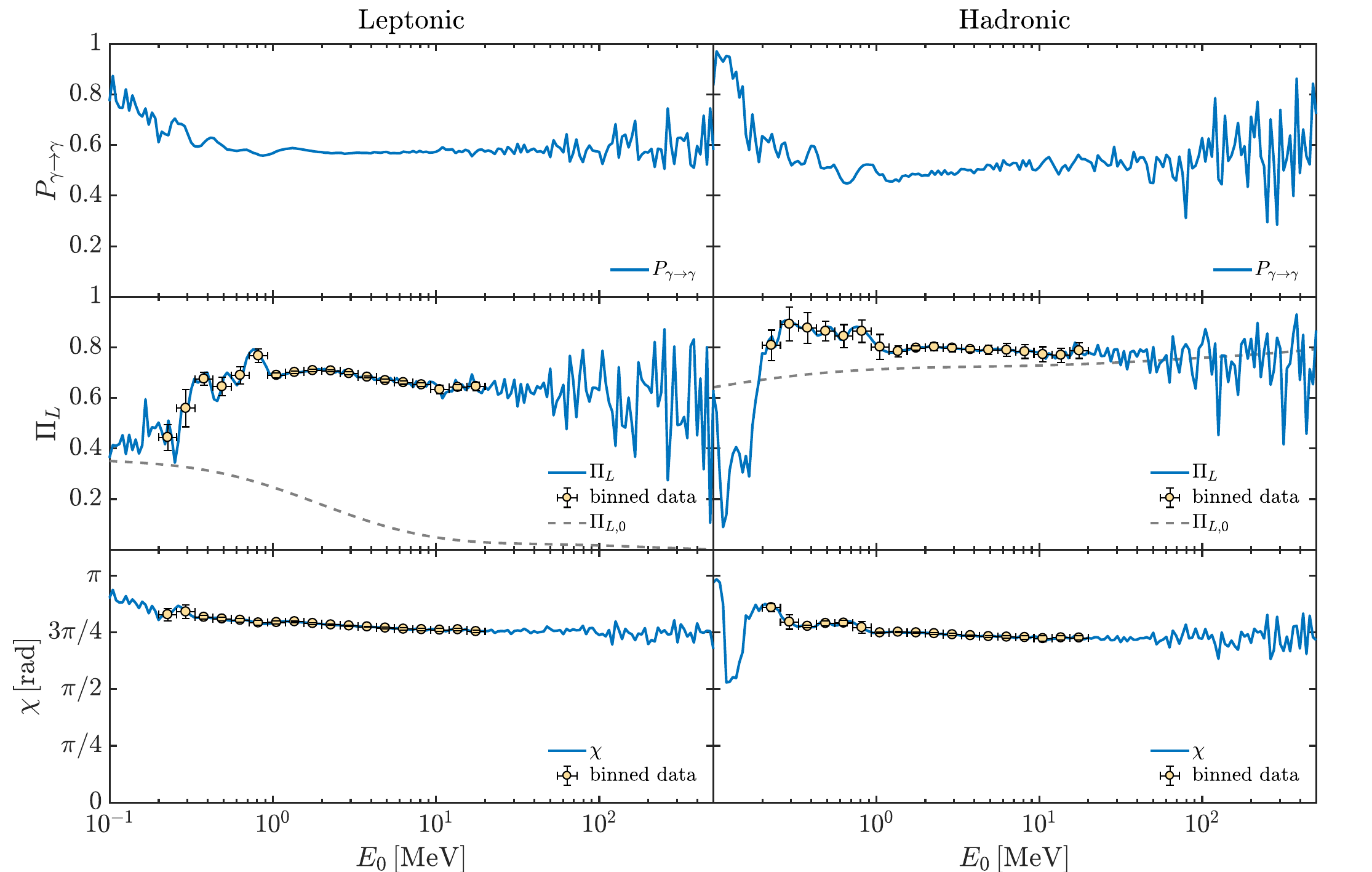}

\includegraphics[width=0.956\textwidth]{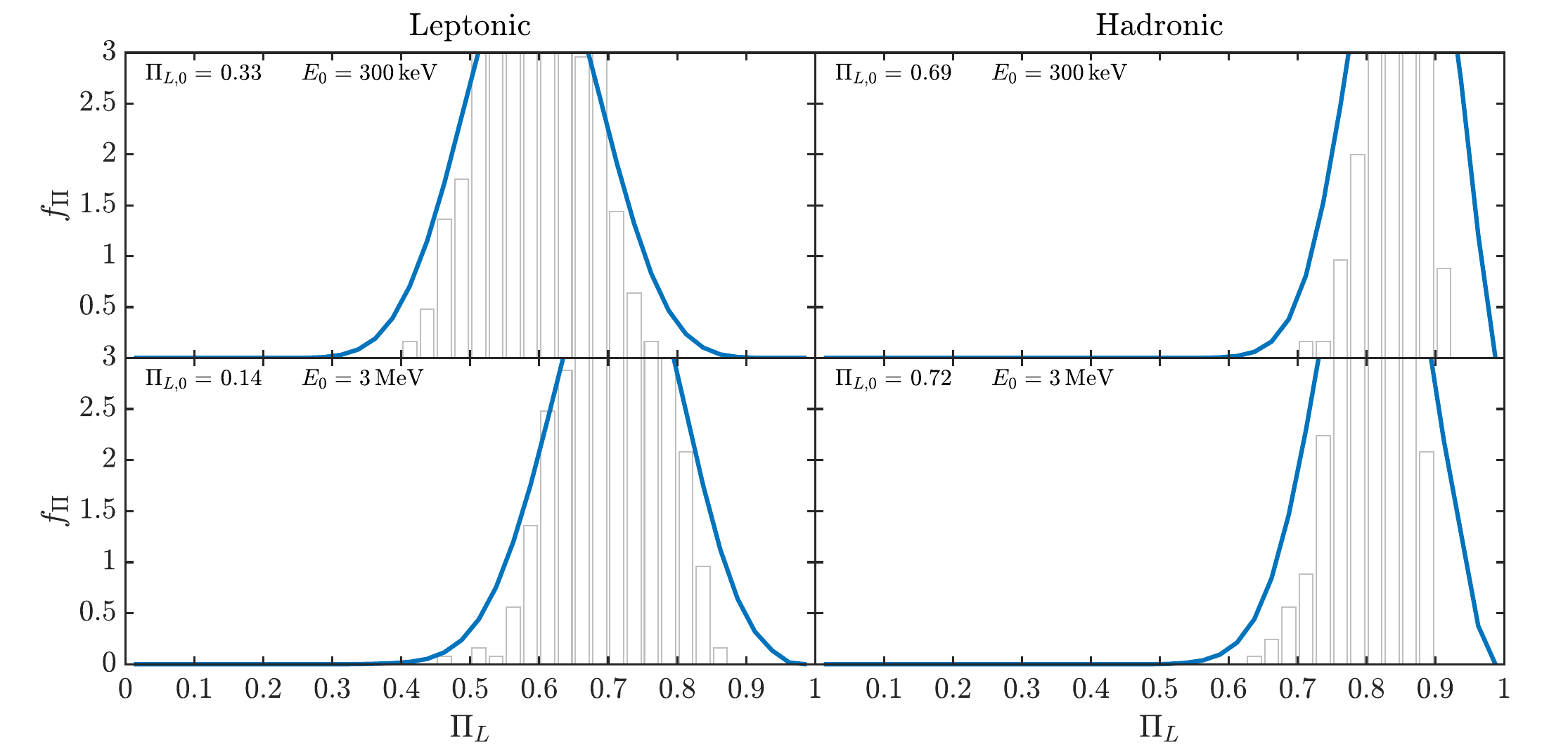}
\caption{OJ~287. Same as Figure~\ref{OJ287HE-14}. We take $g_{a\gamma\gamma}=0.5 \times 10^{-11} \, \rm GeV^{-1}$, $m_a = 10^{-10} \, \rm eV$. The behavior of the initial degree of linear polarization $\Pi_{L,0}$ is shown in Figure~\ref{PiL0}. (Credit:~\cite{grtPolBla}).}\label{OJ287HE-10}
\end{figure}

Concerning the Perseus cluster, we show the behavior of $P_{\gamma \to \gamma}$, $\Pi_L$ and $\chi$ in the top subfigure of Figure~\ref{perseusHE-14} for the case $g_{a\gamma\gamma}=0.5 \times 10^{-11} \, \rm GeV^{-1}$, $m_a \lesssim 10^{-14} \, \rm eV$ and in the top subfigure of {Figure}
~\ref{perseusHE-10} for the case $g_{a\gamma\gamma}=0.5 \times 10^{-11} \, \rm GeV^{-1}$, $m_a = 10^{-10} \, \rm eV$. The associated $f_{\Pi}$ are reported in the bottom subfigures of Figures~\ref{perseusHE-14} and~\ref{perseusHE-10}, respectively. Regarding OJ~287, we likewise report $P_{\gamma \to \gamma}$, $\Pi_L$, and $\chi$ in the top subfigure of Figure~\ref{OJ287HE-14} for the case $g_{a\gamma\gamma}=0.5 \times 10^{-11} \, \rm GeV^{-1}$, $m_a \lesssim 10^{-14} \, \rm eV$ and in the top subfigure of Figure~\ref{OJ287HE-10} for the case $g_{a\gamma\gamma}=0.5 \times 10^{-11} \, \rm GeV^{-1}$,  $m_a = 10^{-10} \, \rm eV$. The corresponding $f_{\Pi}$ are shown in the bottom subfigures of Figures~\ref{OJ287HE-14} and~\ref{OJ287HE-10}, respectively. For OJ~287, we assume (i) the leptonic emission model in the left panels of Figures~\ref{OJ287HE-14} and~\ref{OJ287HE-10} and (ii) the hadronic one in the right panels of Figures~\ref{OJ287HE-14} and~\ref{OJ287HE-10} (see Section \ref{sec4.1} and~\cite{grtPolBla} for more details). Moreover, $f_{\Pi}$ is evaluated for both Perseus and OJ~287 at the two benchmark energies: (i) $E_0 = 300 \, \rm keV$ and \linebreak  (ii) $E_0 = 3 \, \rm MeV$.
\vspace{-6pt}

We first consider the scenario $g_{a\gamma\gamma}=0.5 \times 10^{-11} \, \rm GeV^{-1}$, $m_a \lesssim 10^{-14} \, \rm eV$. In the present situation, the photon--ALP beam propagates in the strong mixing regime, since the photon--ALP mixing term dominates over all other terms inside the mixing matrix of Equation~(\ref{mixmat}) (see Section \ref{sec2}). Consequently, $P_{\gamma \to \gamma}$, $\Pi_L$, and $\chi$ are energy independent in almost all the considered energy band in the case both of photon production in the central region of Perseus, as shown by the top subfigure of Figure~\ref{perseusHE-14}, and of photon generation in the jet of OJ~287, as reported in the top subfigure of Figure~\ref{OJ287HE-14}. We can also infer that, in all the considered cases, the final $\Pi_L$ turns out to be strongly modified with respect to the initial $\Pi_{L,0}$. The binned data show a lower uncertainty than in the UV-X-ray band due to fact that the photon--ALP system lies in the strong mixing regime, which makes both Perseus and OJ~287 good observational targets for missions such as COSI~\cite{cosi}, e-ASTROGAM~\cite{eastrogam1,eastrogam2}, and AMEGO~\cite{amego}. Moreover, regarding OJ~287, we do not observe substantial differences between the leptonic and the hadronic emission model.
\vspace{-6pt}

As far as the behavior of $f_{\Pi}$ is concerned, for both Perseus and OJ~287, the bottom subfigures of Figures~\ref{perseusHE-14} and~\ref{OJ287HE-14} respectively show that the most probable final value of $\Pi_L$ is larger than the initial one $\Pi_{L,0}$ with a substantial broadening. For both Perseus and OJ~287, the most promising energy range to study ALP effects on photon polarization appears around $E_0 = 3 \, \rm MeV$, where the most probable value for $\Pi_L$ is $\Pi_L \gtrsim 0.8$, which is larger than that predicted by conventional physics in all considered cases.


We now move to the scenario $g_{a\gamma\gamma}=0.5 \times 10^{-11} \, \rm GeV^{-1}$, $m_a = 10^{-10} \, \rm eV$. In this situation, the photon--ALP system lies in the weak mixing regime, since the ALP mass term turns out to be non-negligible with respect to the photon--ALP mixing term inside the mixing matrix of Equation~(\ref{mixmat}) (see Section \ref{sec2}). As a result, $P_{\gamma \to \gamma}$, $\Pi_L$, and $\chi$ become energy dependent, as shown in the top subfigure of Figure~\ref{perseusHE-10} regarding Perseus and in the top subfigure of Figure~\ref{OJ287HE-10} concerning OJ~287. Furthermore, we observe that the final $\Pi_L$ turns out to be altered compared with the initial $\Pi_{L,0}$ especially in the case of Perseus and in that of the leptonic model for OJ~287. As already observed in the X-ray band for the case $g_{a\gamma\gamma}=0.5 \times 10^{-11} \, \rm GeV^{-1}$, $m_a \lesssim 10^{-14} \, \rm eV$, the wide energy range, where the photon--ALP system lies in the weak mixing regime, is due to the high variation of the magnetic field and electron number density profiles in the crossed media (see also Sections \ref{sec4.1} and \ref{sec4.3} and~\cite{galantiPol,grtcClu,grtPolBla} for more details). The binned data in the top subfigures of Figures~\ref{perseusHE-10} and~\ref{OJ287HE-10} show that observatories like COSI~\cite{cosi}, e-ASTROGAM~\cite{eastrogam1,eastrogam2}, and AMEGO~\cite{amego} may detect ALP-induced features on $\Pi_L$ for $E_0 \gtrsim 3 \, \rm MeV$ concerning Perseus, and similar features for $E_0 \gtrsim 0.2 \, \rm MeV$ regarding OJ~287 in the case both of the leptonic and of the hadronic model.
\vspace{-3pt}

The behavior of $f_{\Pi}$ reported in the bottom subfigure of Figure~\ref{perseusHE-10} for Perseus and in the bottom subfigure of Figure~\ref{OJ287HE-10} for OJ~287 shows a broadening of the final $\Pi_L$ with respect to the initial $\Pi_{L,0}$ in all the cases. Moreover, the best energy interval for studies of ALP-induced features on $\Pi_L$ appears to be $E_0 \gtrsim 3 \, \rm MeV$ concerning the Perseus cluster, where the most probable value for the final $\Pi_L$ is $\Pi_L \gtrsim 0.2$, while conventional physics predicts $\Pi_L = \Pi_{L,0} = 0$. Even OJ~287 represents a good observational target, since the most probable values of the final $\Pi_L$ are larger than the initial ones, but in this case, conventional physics could justify moderately high values of $\Pi_L$ albeit with some difficulty.

\section{Discussion and Future Perspectives}\label{sec6}

In Section \ref{sec5}, the ALP impact on the polarization of photons coming from Perseus has been reported as an example of ALP effects on galaxy clusters in the X-ray and in the HE band (for details, see~\cite{grtcClu}). Similarly, OJ~287 has been presented as an illustration of the ALP-induced features on photon polarization in BL~Lacs (see~\cite{grtPolBla} for details). Present and planned observatories in the X-ray band~\cite{ixpe,extp,xcalibur,ngxp,xpp} and in the HE range~\cite{cosi,eastrogam1,eastrogam2,amego} possess the capability to detect possible ALP signatures as those analyzed in Section \ref{sec5}, as shown by the realistic binned data reported therein. An analysis about ALP effects on photon polarization in the VHE band has been performed in~\cite{galantiPol}: we will briefly discuss this topic in Section \ref{sec6.1} along with its future crucial importance for the possible identification of an ALP signal.

The ALP-induced features on the final photon degree of linear polarization $\Pi_L$ discussed in Section \ref{sec5} mainly originate inside the galaxy cluster central region and/or in the blazar jet, while the effect of other crossed media is subdominant, as shown \mbox{in~\cite{galantiPol,grtcClu,grtPolBla}}. Specifically, a negligible photon--ALP conversion inside the extragalactic space (when $B_{\rm ext} < 10^{-15} \, \rm G$) does not substantially change the results apart from a smaller broadening of $\Pi_L$ shown in the behavior of its associated probability density function $f_{\Pi}$~\cite{galantiPol,grtcClu,grtPolBla}.

In particular, photons diffusely emitted in the central region of galaxy clusters are expected to be observed as unpolarized within conventional physics both in the X-ray and in the HE range (i.e., with $\Pi_L = 0$, see Section \ref{sec4.3}). Instead, the photon--ALP interaction model consistently predicts $\Pi_L > 0$ for them in both energy ranges. Therefore, galaxy clusters appear among the best observational targets for studies concerning ALP-induced effects on photon polarization. Specifically, a possible future detection of photons diffusely emitted from galaxy clusters with $\Pi_L > 0$ would represent a hint at ALP existence in addition to the already-existing ones~\cite{trgb2012,grdb,gnrtb2023}, since conventional physics along with other scenarios like Lorentz invariance violation~\cite{LIVpol} is unable to explain $\Pi_L > 0$ (see~\cite{grtcClu} for details). BL~Lacs are also promising targets especially in the cases where the photon--ALP interaction model predicts $\Pi_L > 0.8$ (see Section \ref{sec5}), since, in this case, even the hadronic emission model can hardly justify such high polarization degrees within conventional physics (for details, see~\cite{grtPolBla}). However, in case of lower $\Pi_L$ values detected from BL~Lacs, the exclusion of an explanation within conventional physics and the demonstration of the need for ALPs may be less robust. Note that the magnetic field in the jet ${\bf B}^{\rm jet}$ is expected to be quite turbulent, which implies a lower initial photon degree of linear polarization $\Pi_{L,0}$ especially within the leptonic emission model, as shown by a recent theoretical study~\cite{blazarPolNew} and confirmed by IXPE results~\cite{IXPEmrk421,IXPEmrk501,IXPEblLacertae,IXPEblLacertae2,IXPEpg1553,IXPE1ES0229,IXPEmrk421-2,IXPE1ES1959,IXPEPKS2155}. Consequently, the final $\Pi_L$ predicted by the photon--ALP interaction model is reduced in the cases where the initial $\Pi_{L,0}$ remains the most probable final value, as shown by the behavior of the corresponding $f_{\Pi}$ in Section \ref{sec5}.

Other possible targets to evaluate ALP effects on photon polarization are represented by (i) white dwarfs, where ALP bounds can be derived from polarization measurements~\cite{ALPpolMWD,mwd}; (ii) neutron stars, where ALPs produce polarization signatures detectable in the X-ray band~\cite{ALPpol5}; and (iii) GRBs, where the photon--ALP conversion over cosmological distances strongly modifies the initial photon polarization degree, as shown in the pioneering study performed in~\cite{bassan}. We will focus our attention in Section \ref{sec6.2} on GRBs due to their importance also for the Universe transparency problem regarding the recent detection by the LHAASO Collaboration of GRB 221009A up to $18 \, \rm TeV$~\cite{LHAASO,LHAASOspectrumHigh}, which is difficult to justify within conventional physics, but is instead explained by the photon--ALP interaction (see Section \ref{sec6.2} and~\cite{gnrtb2023} for details).

\textls[-25]{As far as the ALP parameter space ($m_a, g_{a\gamma\gamma}$) is concerned, we have analyzed two cases: \linebreak  (i) $m_a \lesssim 10^{-14} \, {\rm eV} , g_{a\gamma\gamma} = 0.5 \times 10^{-11} \, {\rm GeV^{-1}}$ and (ii) $m_a = 10^{-10} \, {\rm eV}$, \mbox{$g_{a\gamma\gamma} = 0.5 \times 10^{-11} \, {\rm GeV}^{-1}$}. Both the previous possibilities are within the CAST bound~\cite{cast}. The former ALP scenario, i.e., the case with $m_a \lesssim 10^{-14} \, {\rm eV}$, is disfavored by some recent bounds~\cite{limFabian,limJulia,limKripp,limRey2}, but this possibility cannot be excluded, as discussed in~\cite{grtPolBla}. Instead, the latter choice, i.e., the case with $m_a = 10^{-10} \, {\rm eV}$, meets all existing bounds and is compatible with the ALP parameters employed in previous studies, where ALP hints have been found~\cite{trgb2012,grdb,gnrtb2023}. This is the reason why the case $m_a = 10^{-10} \, {\rm eV}, g_{a\gamma\gamma} = 0.5 \times 10^{-11} \, {\rm GeV}^{-1}$ appears to be more promising. {As discussed in~\cite{grtcClu}, other possibilities, such as an ALP with mass $m_a = 10^{-11} \, \rm eV$ and coupling to photons $g_{a\gamma\gamma}= 0.5 \times 10^{-11} \, \rm GeV^{-1}$, will produce an intermediate behavior between the two cases considered in Section \ref{sec5} and can meet all current ALP bounds. Note that, while an increment (decrement) of the strength of $g_{a\gamma\gamma}$ implies a corresponding increase (decrease) of the ALP effects on the final $\Pi_L$ and $\chi$ at a given energy, a variation in the value of $m_a$ produces the modification of the energy threshold, starting from which the ALP-induced polarization features become relevant. In particular, as shown in Section \ref{sec5}, for $m_a \lesssim 10^{-14} \, \rm eV$ down to extremely low ALP masses, we do not expect modifications in the ALP-induced effects on $\Pi_L$ and $\chi$, since the ALP mass term turns out to be subdominant with respect to the plasma term in Equation~(\ref{mixmat}). For $10^{-14} \, {\rm eV} \lesssim m_a \lesssim 10^{-10} \, {\rm eV}$, the ALP mass term starts to become important and progressively overcomes the strength of the plasma term with a resulting translation at higher energies of the threshold above which the ALP-induced effects become sizable. For the above-mentioned ALP mass $m_a = 10^{-11} \, \rm eV$, substantial effects on $\Pi_L$ and $\chi$ are expected starting from $E_0 \gtrsim \mathcal{O}(2-5) \, \rm keV$, while, as shown in Section \ref{sec5}, for $m_a = 10^{-10} \, \rm eV$, the ALP impact on $\Pi_L$ and $\chi$ is relevant for $E_0 \gtrsim \mathcal{O}(0.2-0.5) \, \rm MeV$. Lastly, for $m_a$ progressively increasing above $10^{-10} \, \rm eV$, the ALP-induced effects on $\Pi_L$ and $\chi$ will correspondingly appear at larger and larger energies up to disappear in the HE range and to become sizable only in the VHE band and beyond.}

As far as polarization measurements are concerned, unfortunately, we do not presently have many observational data regarding the astrophysical sources considered in this review, i.e., galaxy clusters, blazars, and GRBs. In particular, polarization data in both the X-ray and HE bands regarding diffuse emission from galaxy clusters are substantially missing. Since, as discussed above, these latter astrophysical \textls[-15]{sources are extremely important for analyzing ALP-induced effects on photon polarization and for studying ALP physics in general, a dedicated observational campaign by IXPE~\cite{ixpe} could provide vital information in this regard. Concerning blazars, we now have recent polarimetric data obtained in the X-ray band by IXPE (see, e.g.,~\cite{IXPEmrk421,IXPEmrk501,IXPEblLacertae,IXPEblLacertae2,IXPEpg1553,IXPE1ES0229,IXPEmrk421-2,IXPE1ES1959,IXPEPKS2155}), }which promises to provide us with further important results in the near future. Regarding GRBs, some polarization measurements have been obtained in the X-ray band~\cite{GRBpol1,GRBpol2,GRBpol3,IXPE-GRB221009A}. For all the astrophysical sources discussed in this review, further polarimetric data are expected from new missions proposed both in the X-ray band~\cite{extp,xcalibur,ngxp,xpp} and in the HE range~\cite{cosi,eastrogam1,eastrogam2,amego}.

}

\subsection{Very High Energy Band}\label{sec6.1}

In the VHE band, we consider the case of a BL~Lac emitting within the leptonic model and placed inside a rich galaxy cluster, as studied in~\cite{galantiPol}; therefore, this represents an alternative possibility with respect to those contemplated in Sections \ref{sec5.1} and \ref{sec5.2}. In the present situation, photon--ALP interaction inside the galaxy cluster is relevant, while the conversion inside the host is subdominant (see Section \ref{sec4} for more details). The total transfer matrix of the photon--ALP beam is represented by Equation~(\ref{Utot2}). In the following, two possibilities are evaluated: (i) a BL~Lac with average properties (see below) is located at a redshift $z=0.03$; (ii) the same BL~Lac is placed at $z=0.4$. The BL~Lac redshift represents a crucial parameter to evaluate the photon--ALP system behavior in the VHE band: since, for $E_0 \gtrsim 100 \, \rm GeV$, the absorption of VHE photons due to their interaction with the EBL is not negligible, as it occurs instead for lower energies, as in Sections \ref{sec5.1} and \ref{sec5.2} (see also Section~\ref{sec4.4}), a larger redshift means a higher photon absorption.

In the present scenario, the profiles of the magnetic field in the jet ${\bf B}^{\rm jet}$ and of the electron number density $n_e^{\rm jet}$ are provided by Equations~(\ref{Bjet}) and~(\ref{njet}), respectively, by taking the following parameter values: $B^{\rm jet}_0 = 0.5 \, \rm G$, $y_{\rm em} = 3 \times 10^{16} \, \rm cm$, $n_{e,0}^{\rm jet}=5 \times 10^4 \, \rm cm^{-3}$, and $\gamma = 15$. Photons are assumed to be produced within the leptonic model as unpolarized, i.e., with the initial degree of the linear polarization $\Pi_{L,0}=0$ (see~\cite{galantiPol} for more details). Moreover, we consider a CC galaxy cluster, hosting the BL~Lac, with the electron number density profile
\begin{equation}
\label{eq2}
n_e^{\rm clu}(y)=n_{e,0}^{\rm clu} \left( 1+\frac{y^2}{r_{\rm core}^2} \right)^{-\frac{3}{2}\beta_{\rm clu}},
\end{equation}
with the typical parameter values $n_{e,0}^{\rm clu}=5 \times 10^{-2} \, \rm cm^{-3}$, $r_{\rm core}= 100 \, \rm kpc$, and $\beta_{\rm clu}=2/3$~\cite{cluFeretti,clu2,cluValues}. The cluster magnetic field ${\bf B}^{\rm clu}$ reads from Equation~(\ref{eq1}) by taking the following typical parameter values: $B_0^{\rm clu}=15 \, \upmu{\rm G}$, $\eta_{\rm clu}=0.75$, $q=-11/3$, $k_L=0.2 \, \rm kpc^{-1}$, and $k_H=3 \, \rm kpc^{-1}$~\cite{cluFeretti,clu2,cluValues}. Furthermore, we assume an extragalactic magnetic field strength $B_{\rm ext} < 10^{-15} \, \rm G$, resulting in an inefficient photon--ALP conversion inside the extragalactic space. Concerning the ALP parameters, we take $g_{a\gamma\gamma}=0.5 \times 10^{-11} \, \rm GeV^{-1}$ and $m_a = 10^{-10} \, \rm eV$. 

\begin{figure}[H]
\includegraphics[width=0.75\textwidth]{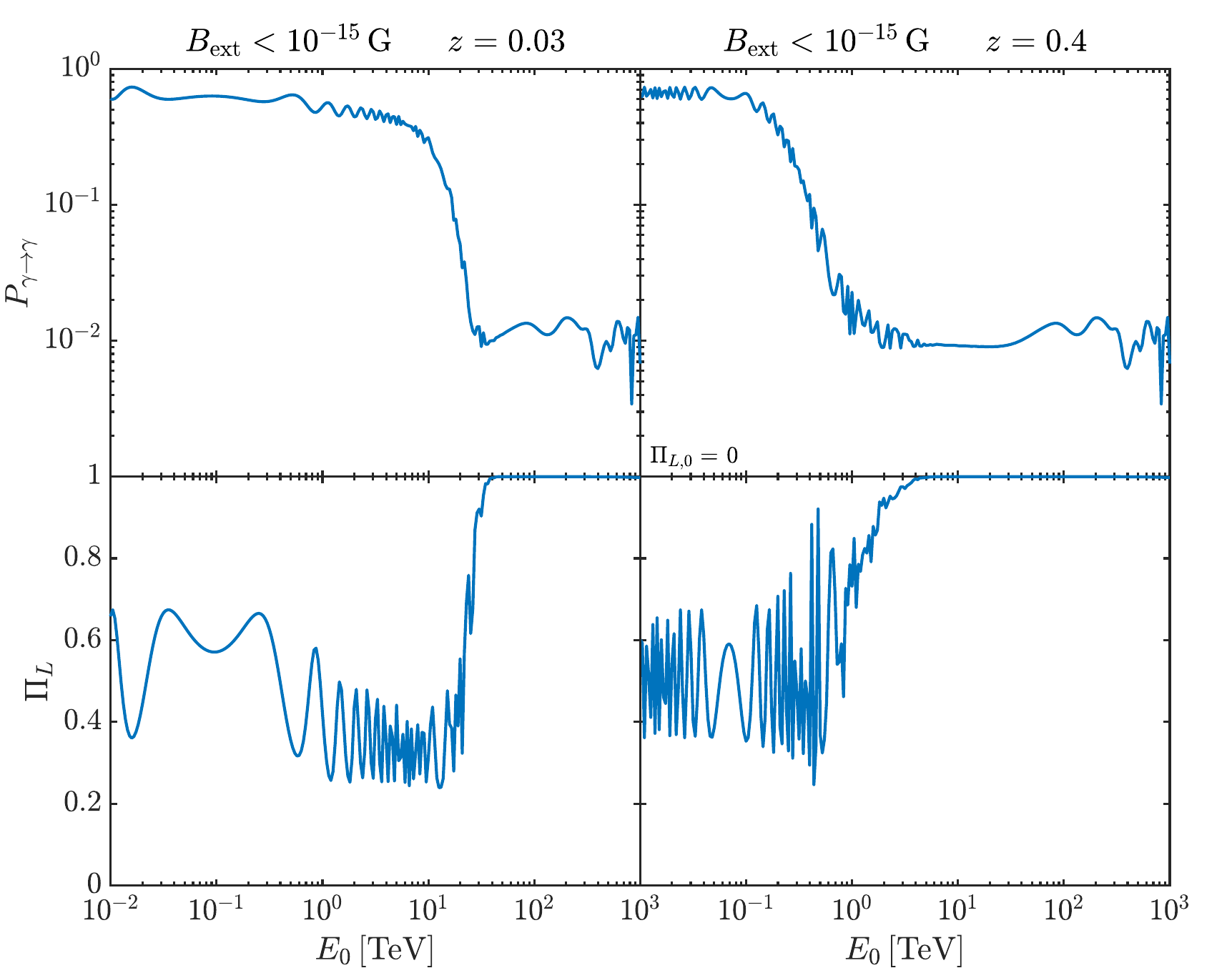}

\includegraphics[width=0.75\textwidth]{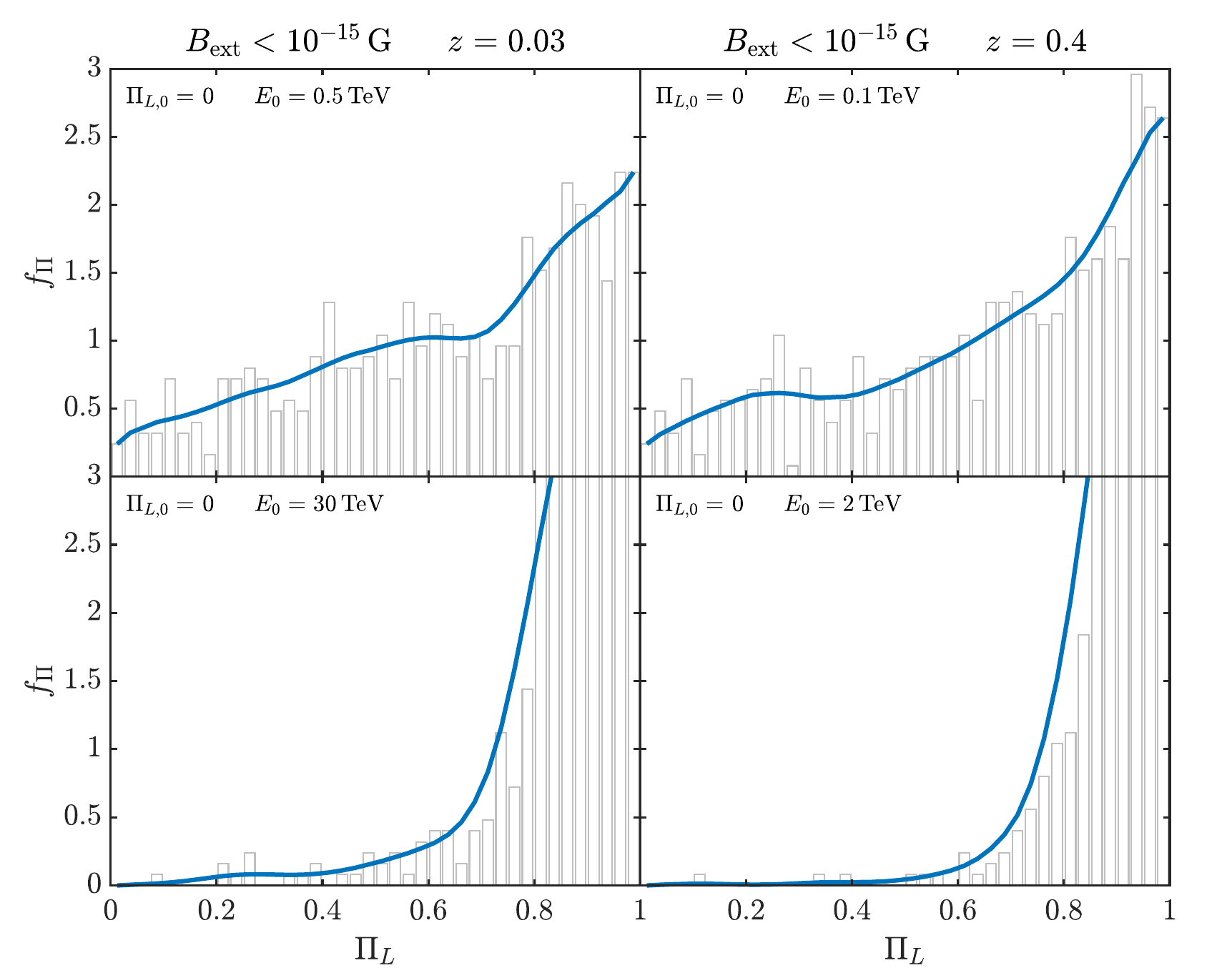}
\caption{BL~Lac emitting within the leptonic model and placed inside a rich galaxy cluster. We consider the  BL~Lac located at a redshift $z=0.03$ or at $z=0.4$ in the left and right columns, respectively. The initial degree of linear polarization is $\Pi_{L,0}=0$. ({\bf Top subfigure}) Photon survival probability $P_{\gamma \to \gamma }$ (upper panels) and corresponding final degree of linear polarization $\Pi_L$ (lower panels) in the energy range $(10^{-2} \text{--} 10^3) \, {\rm TeV}$. We take $g_{a\gamma\gamma}=0.5 \times 10^{-11} \, \rm GeV^{-1}$, $m_a = 10^{-10} \, \rm eV$. ({\bf Bottom subfigure}) Probability density function $f_{\Pi}$ obtained by interpolating the plotted histogram for several realizations of $\Pi_L$ at different energies (see panels). (Credit:~\cite{galantiPol}).}\label{BLLacVHE-10}
\end{figure}

For a BL~Lac with the above-discussed properties, we show $P_{\gamma \to \gamma}$ and $\Pi_L$ in the top subfigure of Figure~\ref{BLLacVHE-10} and the corresponding $f_{\Pi}$ associated with the final $\Pi_L$ in the bottom subfigure. The left panels of Figure~\ref{BLLacVHE-10} report the results for the BL~Lac placed at $z=0.03$, while the right panels show the same results but for the BL~Lac located at $z=0.4$ (see~\cite{galantiPol} for more details). 
\vspace{-6pt}

The top subfigure of Figure~\ref{BLLacVHE-10} shows that $P_{\gamma \to \gamma}$ and $\Pi_L$ are energy dependent, since the photon--ALP system lies in the weak mixing regime due to the strength of the QED one-loop term, which is not negligible at high energies especially inside the BL~Lac jet. Specifically, the final $\Pi_L$ turns out to be strongly modified with respect to the initial one $\Pi_{L,0}$ in both cases of the BL~Lac placed at $z=0.03$ and at $z=0.4$. Now, a new aspect must be considered in the VHE band with respect to the cases at lower energies discussed in Sections \ref{sec5.1} and \ref{sec5.2}: the top subfigure of Figure~\ref{BLLacVHE-10} shows that $P_{\gamma \to \gamma}$ considerably decreases due to the EBL absorption starting from $E_0 \gtrsim 3 \, \rm TeV$ at $z=0.03$ and from $E_0 \gtrsim 200 \, \rm GeV$ at $z=0.4$. Correspondingly, $\Pi_L$ starts to grow up to its limit value $\Pi_L=1$, which is reached at $E_0 \gtrsim 30 \, \rm TeV$ for $z=0.03$ and at $E_0 \gtrsim 5 \, \rm TeV$ for $z=0.4$, where photons are fully polarized and $P_{\gamma \to \gamma} \lesssim 10^{-2}$. The different behavior with respect to the redshift is due to the fact that the EBL absorption grows as both the energy and the redshift increase. As deeply discussed in~\cite{galantiPol}, we have therefore found a peculiar feature: in the presence of photon--ALP interaction, photons, observed at energies where EBL absorption is very strong, are fully polarized. The full explanation of this important feature regarding $\Pi_L$, and shown in the top subfigure of Figure~\ref{BLLacVHE-10}, is provided in~\cite{galantiPol}: in short, after photon--ALP interaction inside the source, we have both photons and ALPs entering the extragalactic space. Then, when EBL absorption is extremely high, all remaining photons are lost during their propagation in the extragalactic space, and we therefore collect only those photons that are reconverted back from ALPs in the Milky Way magnetic field ${\bf B}_{\rm MW}$. Still, only the coherent component of ${\bf B}_{\rm MW}$ is efficient in producing photon--ALP conversion so that photons reconverted back from ALPs inside the Milky Way turn out to be fully polarized.

The behavior of $f_{\Pi}$ in the bottom subfigure of Figure~\ref{BLLacVHE-10} confirms the robustness of the previous discussion. In particular, upper panels deal with the case of small EBL absorption, while the lower ones describe the situation illustrated above, in which $\Pi_L$ tends to its limit value $\Pi_L=1$ with a concomitant high EBL absorption. Note that the behavior of $f_{\Pi}$ in the left panels is similar to that in the right ones, but different energies are considered (see the legend inside the panels) due to the above-described behavior of the EBL absorption as a function of energy and redshift. Since conventional physics cannot explain the scenario $\Pi_L=1$, a possible detection of fully polarized VHE photons not only would be an additional hint at ALPs in addition to those discussed in~\cite{trgb2012,grdb,gnrtb2023}, but also would represent the {\it proof} of ALP existence. Therefore, this ALP-induced feature identified in~\cite{galantiPol} producing fully polarized photons at VHE possesses a high discovery power. However, this possibility nowadays represents only a hope for the future, as currently known techniques cannot measure photon polarization in the VHE band but are limited to tens of GeV at most~\cite{polLimit}.

\subsection{Gamma-Ray Bursts}\label{sec6.2}

ALP effects on the polarization of photons produced by GRBs have been analyzed in a pioneering study~\cite{bassan}, where photon--ALP conversion is evaluated inside the source, in the extragalactic space, inside possibly crossed galaxy clusters, and in the Milky Way within the tens of KeV-few MeV band. In particular, the benchmark energy $E_0 = 100 \, \rm keV$ and the ALP parameters $g_{a\gamma\gamma}= 10^{-11} \, \rm GeV^{-1}$, $m_a = 10^{-13} \, \rm eV$ are taken into account.

GRBs are extremely intense explosions originating at cosmological distances and are characterized by a prompt and an afterglow phase. By considering the typical parameters of the prompt phase---which is analyzed in~\cite{bassan}---regarding the magnetic field strength and the electron number density, the photon--ALP conversion inside the source turns out to be negligible. Regarding the magnetic field structure in extragalactic space, galaxy clusters, and Milky Way, domain-like models with different parameter values inside the several crossed media are assumed in~\cite{bassan}. In particular, by calling $B_{\rm ext}$, $B_{\rm clu}$, and $B_{\rm MW}$ the magnetic field strengths in extragalactic space, galaxy clusters, and Milky Way, respectively, and $L_{\rm dom}^{\rm ext}$, $L_{\rm dom}^{\rm clu}$, and $L_{\rm dom}^{\rm MW}$ the corresponding magnetic field coherence lengths, the following values are assumed: $B_{\rm ext}={\mathcal O}(1) \, \rm nG$, $B_{\rm clu}={\mathcal O}(1) \, \mu \rm G$, $B_{\rm MW}={\mathcal O}(1-5) \, \upmu \rm G$ and \mbox{$L_{\rm dom}^{\rm ext}={\mathcal O}(1) \, \rm Mpc$}, $L_{\rm dom}^{\rm clu}={\mathcal O}(10) \, \rm kpc$, $L_{\rm dom}^{\rm MW}={\mathcal O}(10) \, \rm kpc$ (see~\cite{bassan} for more details).

\begin{figure}[H]
\includegraphics[width=0.67\textwidth]{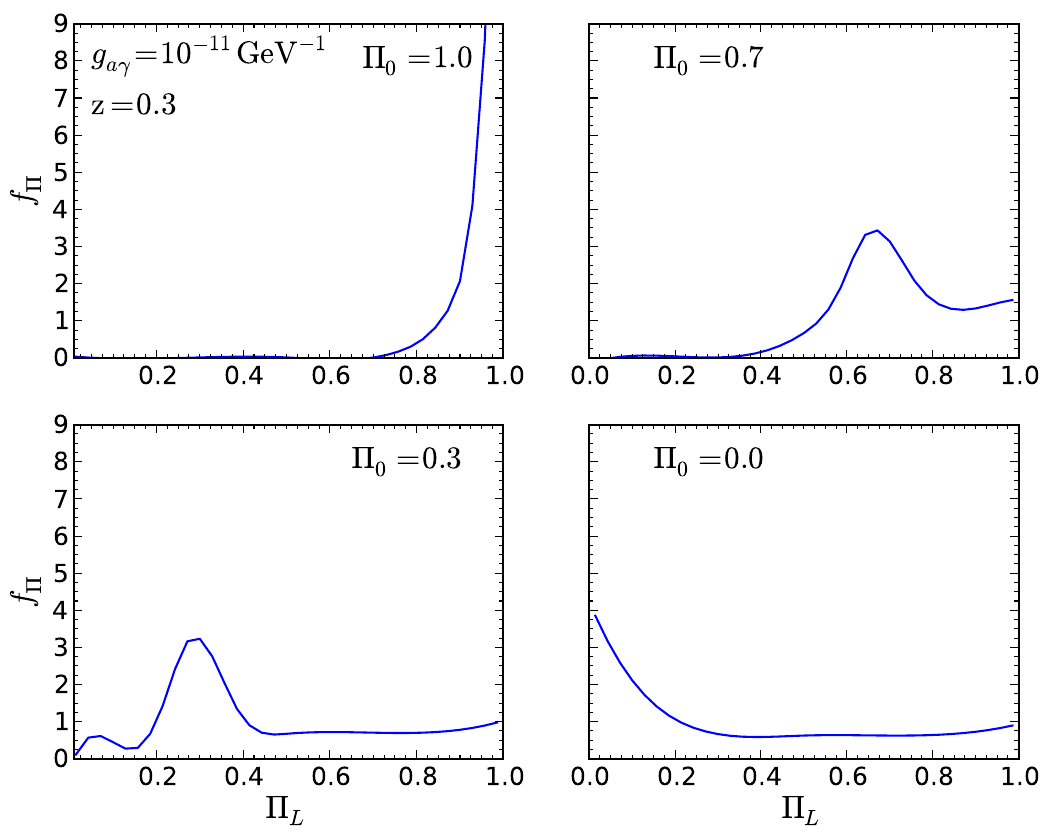}

\includegraphics[width=0.67\textwidth]{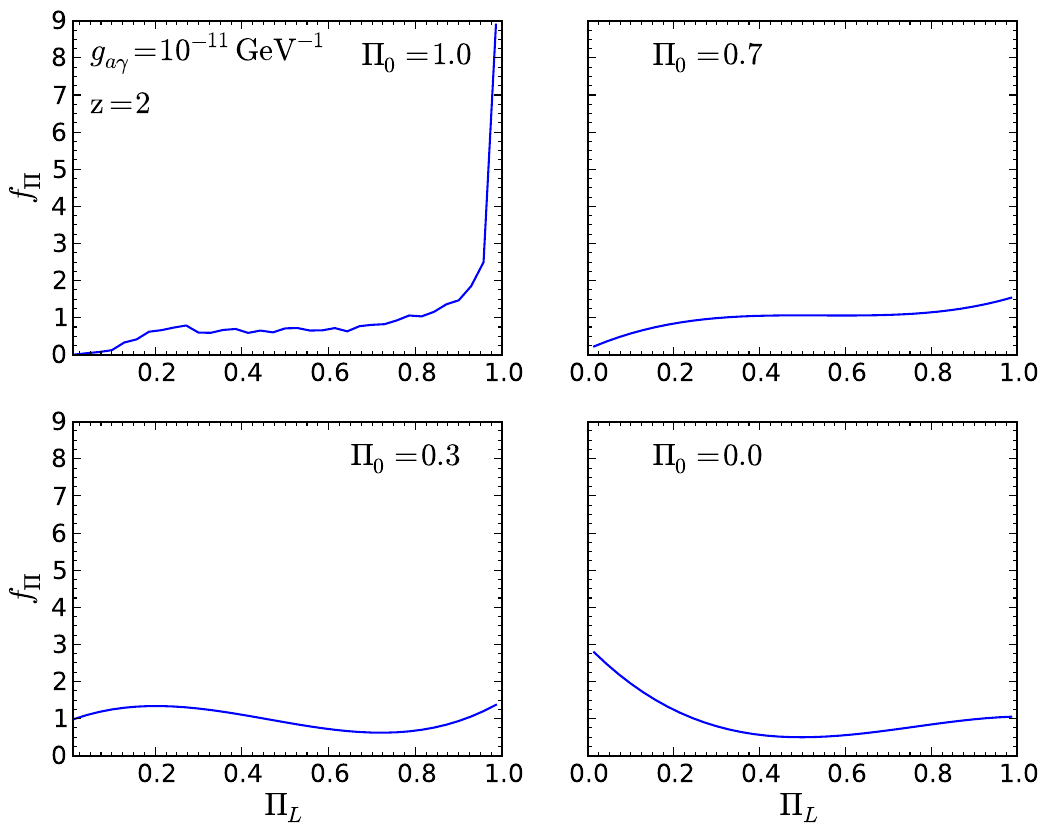}
\caption{GRB probability density function $f_{\Pi}$ for the final photon degree of linear polarization $\Pi_L$ after propagation in the extragalactic space with different values of the initial photon degree of linear polarization $\Pi_0$ (see the legend in the panels). We take $g_{a\gamma}= 10^{-11} \, \rm GeV^{-1}$, $m_a = 10^{-13} \, \rm eV$. ({\bf Top subfigure}) The GRB is placed at redshift $z=0.3$. ({\bf Bottom subfigure}) The GRB is located at $z=2$. (Credit:~\cite{bassan}).}\label{GRBbassan}
\end{figure}

For the chosen parameter values concerning both the photon--ALP system and the crossed media, the photon--ALP beam propagates in the strong mixing regime in all the previous zones within the considered energy band. Consequently, both $P_{\gamma \to \gamma}$ and $\Pi_L$ turn out to be energy independent with $\Pi_L$ which is modified with respect to the initial $\Pi_{L,0}$. Therefore, the only informative quantity in the present situation is represented by $f_{\Pi}$, which is shown in Figure~\ref{GRBbassan} for photon--ALP conversion inside the extragalactic space. In particular, the behavior of $f_{\Pi}$ for different initial values of $\Pi_{L,0}$ is shown in the top subfigure of Figure~\ref{GRBbassan} for a GRB placed at redshift $z=0.3$ and in the bottom subfigure for a GRB located at $z=2$. Note that, in Figure~\ref{GRBbassan}, the initial degree of linear polarization is denoted by $\Pi_0$ and the photon--ALP coupling by $g_{a\gamma}$, but these quantities possess the same meaning of $\Pi_{L,0}$ and $g_{a\gamma\gamma}$, respectively, which are used in the whole review.

Figure~\ref{GRBbassan} shows that the initial $\Pi_{L,0}$ (we recall the different notation here with \mbox{$\Pi_0 \equiv \Pi_{L,0}$}) is strongly modified and broadened in a similar way as it occurs in the systems considered in Sections \ref{sec5.1} and \ref{sec5.2}. In particular, the bottom subfigure shows that this effect increases with the redshift. Moreover, as shown in~\cite{bassan}, the photon--ALP beam propagation inside galaxy clusters and in the Milky Way further increases the broadening effect of the initial $\Pi_{L,0}$ to a point that, at high redshift, the most probable value of the final $\Pi_L$ is no more the initial one, apart from the case of the initial $\Pi_{L,0} \equiv \Pi_0$ = 1.

These findings demonstrate that GRBs are very important astrophysical sources where ALP-induced effects on photon polarization can be studied, 
in addition to galaxy clusters and BL~Lacs (see Sections \ref{sec5.1} and \ref{sec5.2}). The recent detection by the LHAASO Collaboration of photons with energy of up to $18 \, \rm TeV$ produced by GRB~221009A~\cite{LHAASO,LHAASOspectrumHigh}---which is difficult to justify in the context of conventional physics---shows the need to extend the analysis of the ALP impact on GRB polarization even at higher energy bands in a similar way as discussed in Sections \ref{sec4} and \ref{sec5}. Moreover, the explanation of the GRB 221009A detection up to $18 \, \rm TeV$ within the photon--ALP interaction model developed in~\cite{gnrtb2023} demonstrates the importance of the photon--ALP conversion also inside the galaxy hosting the GRB. Therefore, an analysis similar to that performed in~\cite{bassan}, by considering the photon--ALP beam propagation also in the host galaxy and updated models for the photon--ALP conversion inside extragalactic space, galaxy clusters, and Milky Way, appears very promising in order to discover further ALP-induced features on photon polarization and even possible new hints at ALP existence or to improve constraints on the ALP parameter space.

\section{Conclusions}\label{sec7}

In this review, we have described the ALP-induced effects on photon polarization, which can provide a different and complementary way to study the ALP physics with respect to the analysis of the ALP consequences on astrophysical spectra. As mentioned above, there exist three hints at ALP existence linked to the ALP effects on astrophysical spectra, with the most recent and strongest indication coming from the detection by the LHAASO Collaboration of GRB 221009A up to $18 \, \rm TeV$~\cite{LHAASO,LHAASOspectrumHigh}, which can hardly be justified within conventional physics but is naturally explained by photon--ALP interaction~\cite{gnrtb2023} (see also~\cite{ALPinGRB2}). In the near future, we expect that new spectral data from observatories such as ASTRI~\cite{astri}, CTA~\cite{cta}, GAMMA-400~\cite{g400}, HAWC~\cite{hawc}, HERD~\cite{herd}, LHAASO~\cite{LHAASOsens}, and TAIGA-HiSCORE~\cite{desy} will be able to increase our knowledge of the ALP properties.

The study of the ALP effects on the polarization of photons generated by astrophysical sources such as galaxy clusters, blazars, GRBs, neutron stars, and white dwarfs can add further information by shedding light on the ALP parameter space with current spectral hints suggesting $m_a = {\mathcal O}(10^{- 10}) \, {\rm eV}$ and $g_{a \gamma \gamma} = {\mathcal O}(10^{- 11}) \, {\rm GeV}^{- 1}$. As discussed in this review, galaxy clusters represent very promising observational targets in this respect. In fact, according to conventional physics, photons are emitted unpolarized in the central region of galaxy clusters both in the X-ray and in the HE band, resulting in an expected zero polarization degree for the observed photons, while the photon--ALP interaction makes them partially polarized with ALP parameters in the same ballpark of those producing spectral effects. Therefore, a possible future signal of partially polarized photons diffusely produced inside galaxy clusters would be a further strong hint at ALP existence. A negative signal would instead be useful to constrain the ALP parameter space further. In any case, as discussed in this review, future observations of the photon polarization degree from other astrophysical sources like blazars and GRBs are also very promising. Missions such as IXPE~\cite{ixpe} (already operative), eXTP~\cite{extp}, XL-Calibur~\cite{xcalibur}, NGXP~\cite{ngxp}, XPP~\cite{xpp} in the X-ray band and like COSI~\cite{cosi}, e-ASTROGAM~\cite{eastrogam1,eastrogam2}, and AMEGO~\cite{amego} in the HE range may test our findings about ALP effects on photon polarization. 

Finally, an ALP direct detection could be achieved by laboratory experiments such as ALPS II~\cite{alps2}, IAXO~\cite{iaxo} and STAX~\cite{stax}, thanks to the techniques developed by Avignone and collaborators~\cite{avignone1,avignone2,avignone3} and possibly with the ABRACADABRA experiment~\cite{abracadabra}.

\vspace{6pt} 





\funding{The work of the author is supported by a contribution through Grant No. ASI-INAF 2023-17-HH.0 and by the INAF Mini Grant `High-energy astrophysics and axion-like particles', PI: Giorgio Galanti.}

\dataavailability{Data used in this paper can be asked from the authors of the quoted papers.} 




\acknowledgments{The author thanks Giacomo Bonnoli, Silvano Molendi, and Lara Nava for the discussions and all collaborators in this field, especially Enrico Costa, Alessandro De Angelis, Marco Roncadelli, and Fabrizio Tavecchio.}

\conflictsofinterest{The author declares no conflict of interest.}

\begin{adjustwidth}{-\extralength}{0cm}

\reftitle{{References}}

\PublishersNote{}
\end{adjustwidth}
\end{document}